\documentclass[superscriptaddress, prb, a4paper,
showpacs, 
twocolumn,
amsmath,amssymb
]{revtex4}

\usepackage{graphicx}
\usepackage{dcolumn}
\usepackage{bm}

\usepackage{ulem}

\newcommand{\stras}{
	\affiliation{
	Institut de Physique et Chimie des Mat\'eriaux de Strasbourg, UMR~7504 (ULP-CNRS), BP 43,
	F-67034 Strasbourg Cedex 2, France
	}
}
\newcommand{\augs}{
	\affiliation{
	Institut f\"ur Physik, Universit\"at Augsburg, D-86135 Augsburg, Germany
	}
}
\newcommand{\mpipks}{
	\affiliation{
	Max-Planck-Institut f\"{u}r Physik komplexer Systeme, D-01187 Dresden, Germany}
}

\begin{document} 

%\preprint{APS/123-QED}

\title{Lifetime of the first and second collective excitations in metallic nanoparticles}

\author{Guillaume Weick}
\email{gweick@ipcms.u-strasbg.fr}
\stras \augs
\author{Rafael A.~Molina}
\mpipks
\author{Dietmar Weinmann}
\stras
\author{Rodolfo A.~Jalabert}
\stras

\date{\today}

%=============================================================================
%=============================================================================
% ABSTRACT
\begin{abstract}
We determine the lifetime of the surface plasmon in metallic nanoparticles under various conditions,
 concentrating on the Landau damping, which is the dominant mechanism for intermediate-size particles. 
Besides the main contribution to the lifetime, which smoothly increases with the size of the particle, our 
semiclassical evaluation yields an additional oscillating component. For the case of noble metal particles 
embedded in a dielectric medium, it is crucial to consider the details of the electronic confinement; we 
show that in this case the lifetime is determined by the shape of the self-consistent potential near the 
surface. Strong enough perturbations may lead to the second collective excitation of the electronic 
system. We study its lifetime, which is limited by two decay channels: Landau damping and ionization. We 
determine the size dependence of both contributions and show that the second collective excitation
remains as a well-defined resonance.
\end{abstract}

\pacs{78.67.Bf, 73.20.Mf, 71.45.Gm, 31.15.Gy}

%\keywords{Collective excitations; Optical properties of clusters; Exchange, correlation, dielectric and magnetic
%response functions, plasmons; Semiclassical methods}

\maketitle

%\tableofcontents

%=============================================================================
%=============================================================================
%=============================================================================
%=============================================================================
% INTRODUCTION
\section{Introduction}
The surface plasmon (SP) resonance is a very important collective excitation in metallic clusters. 
\cite{kreibig,deheer} It is the dipolar vibration of the electronic center of mass with respect to the positive
ionic charge, analogous to the giant resonance of nuclei. \cite{bertsch} Since an external electromagnetic
 dipole field couples directly to the electronic center of mass, the photoabsorption spectrum
of a metallic cluster is dominated by the SP. The lifetime of this collective excitation is a determining
factor in the relaxation process studied in femtosecond pump-probe experiments.\cite{bigot} 

The classical electromagnetic theory for a charged metallic sphere in the vacuum yields the energy 
$\hbar \omega_{\rm M}$ of the resonance, with the Mie frequency
$\omega_{\rm M}=\omega_{\rm p}/\sqrt{3}$, where $\omega_{\rm p} = (4\pi n e^2/m_{\rm e})^{1/2}$ is the bulk
plasma frequency of the metal, $n$, $e$, and $m_{\rm e}$ being the electron density, charge, and
mass, respectively. If the clusters are embedded in a matrix (of dielectric constant $\epsilon_{\rm m}$)
and/or we consider noble metal  clusters (where the effect of the d electrons can be mo\-deled by a
dielectric function $\epsilon_{\rm d}$) the Mie frequency takes the form 
$\omega_{\rm M} = \omega_{\rm p}/\sqrt{\epsilon_{\rm d}+2\epsilon_{\rm m}}$. The Mie
frequency is close to the experimentally measured resonances. Such an agreement is not surprising 
since we deal with a collective excitation with a clear classical counterpart. Small red- and blueshifts with
respect to $\omega_{\rm M}$ have been experimentally observed in different phy\-si\-cal conditions, and 
various microscopic approaches have been developed to account for the frequency shifts. 
\cite{kreibig,deheer} The most successful among them are based on a jellium description (where 
the ionic positive charges are taken as a uniform background) and linear-response theory in the framework  
of the time-dependent local density approximation (TDLDA). \cite{ekardt} 

While it is difficult to measure the SP lifetime, numerous data for the linewidths of the absorption peak
of ensembles of nanoparticles are available, \cite{kreibig, deheer, brechignac, charle}
but their theoretical analysis has proven to be quite involved. In principle, inhomogeneous effects arising
from the dispersion among the probed ensemble of clusters have to be separated from the pro\-per\-ties of
single particles. Intrinsic effects depending on the bulk properties of the metal have to be separated from
size-dependent properties of the cluster, and from the effect of the interaction with the local environment 
(matrix). In addition, the decay of the SP may follow different channels depending on the size of the 
cluster. \cite{deheer} We calculate the Landau damping contribution to the linewidth
(i.e.~decay into particle-hole pairs), which dominates in the case of small clusters of radius $a$ in the range
0.5--2.5 nm. \cite{bertsch} For larger particles the Landau damping competes with radiation damping.

Recent measurements of single-cluster optical absorption have rendered accessible the optical properties 
of individual nano-objects. \cite{klar, lamprecht, arbouet} Most of the individual nanoparticles studied so
far (in static \cite{klar} or dynamic \cite{lamprecht} setups) are too large to be in the Landau regime.
However the linewidth of a single $2.5$ nm radius gold nanoparticle has been determined lately. \cite{arbouet}
The possibility of overcoming the inhomogeneous broadening, and the application as biological markers,
resulted in a renewed interest for the optical response of metallic clusters.

Kawabata and Kubo studied the Landau damping of the SP,\cite{kubo} and using linear
response theory they proposed a total linewidth 
\begin{equation*}
\Gamma_{\rm t}(a) = \Gamma_{\rm i} + \Gamma(a) ,
\end{equation*}
with $\Gamma_{\rm i}$ a constant intrinsic width and $\Gamma(a)$ inversely proportional to the particle
size $a$. Barma and Subrahmanyam, \cite{barma} and Yannouleas and Broglia \cite{yannouleas}
improved this calculation and proposed corrections to the behavior of $\Gamma$ outside the regime
$\hbar \omega_{\rm M}/ \varepsilon_{\rm F} \ll 1$, where $\varepsilon_{\rm F}$ is the Fermi energy. They
obtained
\begin{equation}
\label{sp_smooth}
\Gamma (a)  = \frac 32 \frac{\varepsilon_{\rm F}}{k_{\rm F}a} g(\xi) ,
\end{equation}
where $k_{\rm F}$ is the Fermi wave-vector, and $g$ a function of the ratio 
$\xi=\hbar\omega_{\rm M}/\varepsilon_{\rm F}$. Numerical calculations within the TDLDA on free alkaline 
clusters \cite{yavibro} agree with this analytical result for $1.5 \leqslant a \leqslant 2.5$ nm. For smaller 
radii, $\Gamma$ shows a nonmonotonous size dependence superposed to the overall behavior
of Eq.~\eqref{sp_smooth}. These shell effects arise from the electron-hole density-density
correlations in the angular-momentum restricted density of states, \cite{molina} and a semiclassical 
evaluation of $\Gamma$ is in good agreement with TDLDA calculations.

The experimentally observed nonmonotonous size dependence of the plasmon linewidth for charged alkaline
metal nanoparticles in vacuum \cite{brechignac} is consistent with the theoretical calculations. However,
our calculated linewidths yield lower bounds for the experimentally measured ones, and the corresponding
lifetimes are upper bounds of those found in real systems. Further measurements with smaller radii and a narrower
size-distribution seem necessary to clearly establish the connection with the theory.

Noble metal clusters embedded in inert matrices \cite{charle} also exhibit a nonmonotonous
linewidth for small $a$. However, a direct application of Eq.~\eqref{sp_smooth} overestimates the smooth 
part of $\Gamma$. \cite{molina} This discrepancy motivates us to develop a refined theoretical description  
of the SP lifetime for the case of clusters with internal dielectric constant $\epsilon_{\rm d}$ and 
surrounded by a dielectric medium with constant $\epsilon_{\rm m}$. The presence of an inhomogeneous 
dielectric environment leads to the modification of Eq.~\eqref{sp_smooth}.

Under a weak initial optical excitation, only the first surface plasmon (that we simply denote ``surface 
plasmon" when there is no possibility of confusion) is excited. With sufficiently strong initial excitations, 
we can also reach the second quantum level of the center-of-mass motion, known as the second (or 
double) plasmon. Such a resonance will be experimentally relevant provided its lifetime is sufficiently 
large (in the scale of $\omega_{\rm M}^{-1}$). The lifetime is given by the anharmonicities of the
center-of-mass system and by its interactions with the other degrees of freedom. Like in the previous 
discussion, the Landau damping is an important channel for the decay of the se\-cond plasmon, but a 
new 
channel appears when $2 \hbar \omega_{\rm M}$ is larger than the ionization energy: the ionization in 
which the cluster looses an electron into the continuum. \cite{bertsch:dp} Such a process was discussed 
in order to interpret the ionization of charged Na$_{93}^+$ clusters observed by Schlipper and 
collaborators. \cite{schlipper}

We calculate the decay rates associated with different channels for the single- and double-plasmon states
using a semiclassical approach within a mean-field description of the nanoparticle. Whenever it is 
possible, we verify the semiclassical  approach by comparing to numerical calculations. We characterize
the size-dependent oscillations of the first and second plasmon linewidths for the case of free alkaline 
metals. In addition, we analyze the the\-o\-re\-ti\-cal difficulties in extending these calculations to the case of 
embedded and/or noble metal clusters and propose a way to overcome them. We also apply our 
semiclassical approach to the calculation of the ionization rate via the double-plasmon channel, and 
obtain results comparable with the experiments. \cite{schlipper}

The paper is organized as follows: In Sec.~\ref{sec_PPA} we introduce the basic formalism
for the photoabsorption and the SP linewidth. In Sec.~\ref{sec_sp} we present the
semiclassical calculation of the single-plasmon linewidth, testing some of the approximations that we will 
use in the sequel. In Sec.~\ref{sec_inhomogeneous} we study the case of noble metal nanoparticles 
embedded in a dielectric medium and present the need to improve the existing theory for this case. 
In Sec.~\ref{sec_dp} we show a semiclassical description of the two main channels contributing to the
decay of the double plasmon: Landau damping and ionization. Finally in Sec.~\ref{sec_ccl} we draw
the conclusions and the perspectives of our work. We relegate to the appendix a few technical, but
important issues; in Appendix \ref{mismatch} we extend the standard calculation of the plasmon linewidth
to the case where the cluster is made of a noble metal and/or is embedded in a nonreactive matrix.
In  Appendix \ref{sec:semiclassics} we show how to take advantage of the spherical symmetry in semiclassical
calculations like the ones of this paper, and how to recover some well-known results. In 
Appendix \ref{sec_gh} we present the frequency dependence of the different plasmon linewidths.

%=============================================================================
%=============================================================================
%=============================================================================
%=============================================================================
% PHOTABSORPTION AND PLASMON-POLE APPROXIMATION
\section{Photoabsorption and plasmon linewidth}
\label{sec_PPA}
When the cluster is placed in an electromagnetic field with a wavelength much larger than its size, 
\cite{wavelength} the photoabsorption cross section is obtained from the application of the dipole
operator on the ground state of the system:
\begin{equation}
\label{cross-section}
\sigma(\omega) = \frac{4\pi e^2 \omega}{3c} \sum_f \left|\langle f|z|0\rangle \right|^2
\delta(\hbar\omega-E_f) ,
\end{equation}
where $c$ is the velocity of light; $| f \rangle$ and $E_f$ are, respectively, the many-body excited states
and eigenenergies of the electronic system. The ground state is noted as $| 0 \rangle$ and its energy is
taken as zero. In Eq.~\eqref{cross-section}, the photon degrees of freedom have already been integrated 
out. In order to describe the electronic system, we consider a closed shell nanoparticle (perfectly spherical 
with a ``magic number" of atoms) within a jellium model. The Hamiltonian re\-pre\-sen\-ting $N$ valence 
electrons in a uniformly positively charged sphere of charge $+Ne$ is given by 
\begin{equation}
\label{H}
H = \sum_{i = 1}^N \left[ \frac{p_i^2}{2 m_{\rm e}} + U(r_i) \right] + \frac{1}{2} 
\sum_{\substack{i, j = 1\\(i \neq j)}}^N 
\frac{e^2}{\left| {\bf r}_i - {\bf r}_j \right|} ,
\end{equation}
with the single-particle confining potential 
\begin{equation}
\label{harlomb}
U(r) = 
\begin{cases}
	\displaystyle 2 \pi n e^2 \left( \frac{r^2}{3}-a^2 \right)  , &   r \leqslant a  ,  \\
	\displaystyle -4 \pi n e^2 \frac{a^3}{3r}  , &   r > a  ,
\end{cases}
\end{equation}
where $n = N/\cal V$ is the electronic density and ${\cal V} = 4 \pi a^3/3$ the volume of the particle. The 
potential $U$ is harmonic inside the particle and Coulomb-like outside it.

The experimentally obtained photoabsorption cross section is dominated by the surface plasmon (SP)
re\-so\-nan\-ce at the frequency $\omega_{\rm M}$. The width $\Gamma$ of this resonance can, in principle, 
be calculated from the eigenstates of $H$ appearing in Eq.~\eqref{cross-section}. However, this 
procedure is in general exceedingly difficult, and thus various approximation schemes have been 
proposed. \cite{deheer, yannouleas} Among them, the TDLDA (time-dependent local density 
approximation) is a numerical approach based on the local density approximation. \cite{ekardt} We will 
use this numerical scheme as a check of analytical approaches that instructs us on the physical 
underlying mechanisms to the plasmon decay. 

Since we are in the long-wavelength limit where the field couples only to the electronic center of mass, a
particularly useful decomposition of the Hamiltonian \eqref{H} is 
\begin{equation*}
H = H_{\rm cm}+H_{\rm rel}+H_{\rm c} .
\end{equation*}
Introducing the canonical coordinates $({\bf R}, {\bf P})$, the (harmonic) center-of-mass Hamiltonian is 
given by
\begin{equation*}
H_{\rm cm} = \frac{P^2}{2Nm_{\rm e}}+\frac 12 Nm_{\rm e} \omega_{\rm M}^2 R^2  .
\end{equation*}
$H_{\rm rel}$ is the Hamiltonian of the relative coordinates and $H_{\rm c}$ expresses the coupling
between the two subsystems. Introducing the standard position, momentum and lowering operators 
\begin{equation*}
a_Q = \sqrt{\frac{Nm_{\rm e} \omega_{\rm M}}{2 \hbar}} Q+\frac{\rm i}{\sqrt{2 Nm_{\rm e} \hbar \omega_{\rm M}}}P_Q 
 , \ Q=X, Y, Z,
\end{equation*}
we can write 
\begin{equation*}
\hat H_{\rm cm} =  
\hbar \omega_{\rm M} \sum_{Q=X, Y, Z} \left( {a}_{Q}^{\dagger} {a}_{Q} +\frac 12 \right) .
\end{equation*}

It is difficult to handle $H_{\rm rel}$ and $H_{\rm c}$ in the general case. A notable exception is that
of a  confining potential which is not given by Eq.~\eqref{harlomb}, but which is harmonic  for all $r$. We 
are then in the situation for which Kohn's theorem \cite{kohn} applies. It states that in a purely harmonic 
confinement potential, and with interactions only depending on the interparticle distance, the 
center of mass and the relative coordinates decouple (i.e., $H_{\rm c}=0$). The motion of the
center of mass is that of a harmonic oscillator, with the characteristic frequency of the confining
potential, independent of the electron-electron interaction. Due to the decoupling, the SP has an infinite 
lifetime. Kohn's theorem gives us a first insight into the relaxation process of the SP: The Coulomb part of
$U$ (for $r > a$) leads to the coupling of the center of mass and the relative coordinates
(i.e.~$H_{\rm c} \neq 0$), and translates into the decay of the SP.

For the realistic situation $H_{\rm c} \neq 0$, it is useful to describe $H_{\rm rel}$ and $H_{\rm c}$
within the mean-field approximation, where $H_{\rm rel}$ can be expressed as
\begin{equation*}
\hat H_{\rm MF} = \sum_{\alpha} \varepsilon_{\alpha} c_{\alpha}^{\dagger} c_{\alpha} ,
\end{equation*}
where $\varepsilon_{\alpha}$ are the eigenenergies for the mean-field potential $V$ and 
$c_{\alpha}^{\dagger}$ ($c_{\alpha}$) creates (annihilates) the one-body eigenstate $| \alpha \rangle$. 
Consequently, the mean-field approximation to $H_{\rm c}$ will be given by the change $\delta V$ 
induced in the one-body potential $V$ by a displacement $Z$ of the center of mass. In
Appendix \ref{mismatch} we show how to obtain $\delta V$ in a self-consistent way from the electronic
Coulomb interactions [Eq.~\eqref{deltaV}]. In second quantization, we can write
\begin{equation}
\label{Hc}
\hat H_{\rm c} = \sqrt{\frac{\hbar m_{\rm e} \omega_{\rm M}^3}{2 N}} 
\sum_{\alpha \beta} d_{\alpha \beta} ( a_Z^{\dagger} + a_Z ) c_{\alpha}^{\dagger} c_{\beta} ,
\end{equation}
where $d_{\alpha \beta}$ is the matrix element of the classical dipole field between two eigenstates of the 
unperturbed mean-field problem [Eq.~\eqref{dipole_field}].

The laser excitation induces an initial electronic state corresponding to a rigid shift (with a magnitude $Z$) 
of the unperturbed ground state. Within our separation for the degrees of freedom of the electronic
system, such an initial state can be written as a product of the ground state for the relative coordinate
system and a coherent state for the center of mass (along the direction of the excitation). Since the
amplitude of the perturbation is assumed to be small, the initial coherent state can be  approximated by a 
linear superposition of the ground state $| 0_{\rm cm} \rangle$ and the first (harmonic oscillator) excited
state $| 1_{{\rm cm}, Z} \rangle$. The lifetime of such an initial state is that of the SP. It is related to the
transition rate $\Gamma$ of $| 1_{{\rm cm}, Z} \rangle$ to $| 0_{\rm cm} \rangle$ by
$T_1 = \hbar/\Gamma$,  while the dephasing time is given by $T_2 = 2 T_1$. This decay is due to the coupling
$\hat H_{\rm c}$ and results in the transition of the relative coordinate system from its ground state to
excited ones (that within our mean-field assumption we note $| 0_{\rm MF} \rangle $ and
$| f_{\rm MF} \rangle$, respectively).

Assuming a weak coupling $\hat H_{\rm c}$, the SP linewidth can be obtained from the Fermi Golden 
Rule as
\begin{equation*}
\Gamma = 2 \pi \sum_{f_{\rm MF}}
\left| \langle 0_{\rm cm}, f_{\rm MF} | \hat H_{\rm c} | 1_{{\rm cm}, Z},  0_{\rm MF} \rangle \right|^2 
\delta(\hbar \omega_{\rm M}-\varepsilon_{f_{\rm MF}}) .
\end{equation*}
According to form \eqref{Hc} of $\hat H_{\rm c}$, the final mean-field states $| f_{\rm MF} \rangle$
are particle-hole excitations, and therefore
\begin{equation}
\label{Gamma_sp}
\Gamma  = \frac{\pi \hbar \omega_{\rm M}^3 m_{\rm e}}{N} \sum_{ph}  \left| d_{ph} \right|^2 
\delta(\hbar \omega_{\rm M} - \varepsilon_p+\varepsilon_h) ,
\end{equation}
where $| p \rangle$ and $| h \rangle$ represent, respectively, particle and hole states of the mean-field 
problem. 

Form \eqref{Gamma_sp} of the SP linewidth can also be derived from discrete-matrix random phase
approximation \cite{yannouleas} using the classical field associated with the collective state as the source 
of particle-hole transitions. The procedure presented above is easy to generalize for the two cases 
important for our work: a nonhomogeneous dielectric function and the excitation of the second plasmon.

%=============================================================================
%=============================================================================
%=============================================================================
%=============================================================================
%=============================================================================
%=============================================================================
%=============================================================================
%=============================================================================
\section{Size dependence of the plasmon linewidth}
\label{sec_sp}
\begin{figure}[t]
\begin{center}
\includegraphics[width=8truecm]{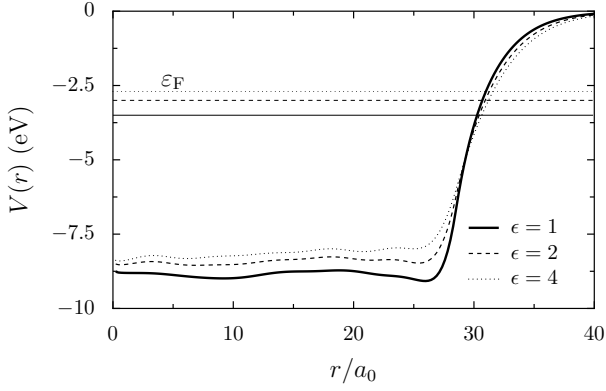}
\caption{\label{fig_slope} Self-consistent potential as a function of the radial coordinate (in units of the 
Bohr radius $a_0$) from the TDLDA calculations for a $832$-atom nanoparticle with mean distance 
between electrons $r_{\rm s} = (3/4 \pi n)^{1/3} = 3.03$ $a_0$, corresponding to $a \simeq 28.5$ $a_0$. 
The different curves are for $\epsilon=\epsilon_{\rm d}=\epsilon_{\rm m}$ between $1$ and $4$, showing 
that the slope of the potential decreases with increasing values of $\epsilon$. The corresponding
Fermi levels are indicated by horizontal lines.}
\end{center}
\end{figure}

In order to evaluate the plasmon linewidth from Eq.~\eqref{Gamma_sp}, we need a description of the
eigenstates $| \alpha \rangle$ ($p$ or $h$) of the mean-field problem. The self-consistent potential
obtained from TDLDA (thick line, Fig.~\ref{fig_slope})
suggests that for analytical calculations, $V(r)$ can 
be approximated by a spherical well of radius $a$ and finite height $V_0$:
\begin{equation}
\label{staircase}
V(r) = V_0 \Theta(a-r) ,
\end{equation}
with $\Theta$ the Heaviside distribution. This stair like ap\-pro\-xi\-ma\-tion becomes more appropriate
as the particle size increases. As we discuss in the next chapter, the dielectric constants inside and 
outside the cluster influence the steepness of $V(r)$.

The spherical symmetry of the problem allows us to separate the wave functions and matrix elements into 
radial and angular components
\begin{equation*}
\psi_{klm}({\bf r}) = \frac{u_{kl}(r)}{r} Y_{l}^{m}(\Omega) ,
\end{equation*} 
and 
\begin{equation}
\label{dab}
d_{\alpha \beta} =
{\cal A}_{l_\alpha l_\beta}^{m_\alpha m_\beta} {\cal R}_{k_\alpha k_\beta}^{l_\alpha l_\beta} .
\end{equation}
$u_{kl}$ satisfies the radial Schr\"odinger equation \eqref{1DSch}, $Y_{l}^{m}$ represents the spherical 
harmonics, $k = (2 m_{\rm e} \varepsilon)^{1/2}/\hbar$ is given by the principal quantum number, while 
$l$ and $m$ are the angular momentum quantum numbers. The angular part of the matrix element can be 
expressed in terms of the Wigner-$3j$ symbols as
\begin{align}
\label{sp_angular}
{\cal A}_{l_\alpha l_\beta}^{m_\alpha m_\beta} =
&(-1)^{m_\alpha} \sqrt{(2l_\alpha+1)(2l_\beta+1)} 
\nonumber \\
&\times
\begin{pmatrix}
l_\alpha & l_\beta & 1 \\ 
0 & 0 & 0
\end{pmatrix}
\begin{pmatrix}
l_\alpha & l_\beta & 1 \\ 
-m_\alpha & m_\beta & 0
\end{pmatrix}
.
\end{align}
The dipole matrix element of the radial problem can be written as
\begin{equation}
\label{slope_dipole}
{\cal R}_{k_\alpha k_\beta}^{l_\alpha l_\beta} = 
\frac{\hbar^2}{m_{\rm e} (\varepsilon_\alpha -\varepsilon_\beta)^2} 
\int_0^\infty {\rm d}r \ u_{k_\alpha l_\alpha}^\ast(r) \ \frac{{\rm d}V}{{\rm d}r} \ u_{k_\beta l_\beta}(r)  .
\end{equation}
In the limit of large $V_0$, we have $u_{kl}(r) = \sqrt{2} [a^{3/2} j_{l+1}(ka)]^{-1} r  \, j_l(kr)$, where $j_l$ are 
the spherical Bessel functions and the allowed values of $k$ are given by the quantization condition 
$j_{l}(ka) = 0$, one obtains \cite{yannouleas}
\begin{equation}
\label{sp_radial}
{\cal R}_{k_pk_h}^{l_pl_h} = \frac{2 \hbar^2}{m_{\rm e} a} \frac{\sqrt{\varepsilon_p \varepsilon_h}}
{{(\varepsilon_p-\varepsilon_h)}^2}  .
\end{equation}

The summations appearing in Eq.~\eqref{Gamma_sp} can be replaced by integrals provided one knows 
the particle (and hole) density of states (DOS). Decomposing the latter as a sum over its fixed angular 
momentum components 
[$\varrho(\varepsilon) = \sum_{l = 0}^\infty \sum_{m = -l}^{+l} \varrho_l(\varepsilon)$], we have
\begin{align*}
\Gamma = 2
 \frac{4 \pi \hbar}{Nm_{\rm e} a^2 \omega_{\rm M}} 
&\int_{\max{(\varepsilon_{\rm F}, \hbar \omega_{\rm M})}}^{\varepsilon_{\rm F} + \hbar \omega_{\rm M}}
{\rm d}\varepsilon_p \ 
\varepsilon_p \varepsilon_h 
\\
&\times
\sum_{\substack{l_p, m_p\\l_h, m_h}} \varrho_{l_p}(\varepsilon_p) 
\varrho_{l_h}(\varepsilon_h) \left( {\cal A}_{l_pl_h}^{m_pm_h} \right)^2  ,
\nonumber
\end{align*}
with $\varepsilon_h = \varepsilon_p-\hbar \omega_{\rm M}$.
The overall factor of $2$ accounts for the spin degeneracy. The angular part \eqref{sp_angular} of the 
dipole matrix element contains the selection rules $m_h=m_p$ and $l_h=l_p \pm 1$. Performing the sum 
over $m_p$ and $l_h$, with the change of variables  $\varepsilon_p = \varepsilon_0 \eta_p^2$, 
$\varepsilon_h =  \varepsilon_0 \eta_h^2$ ($\varepsilon_0 = \hbar^2/2 m_{\rm e} a^2$ and $\eta = ka$), 
we have 
\begin{align}
\label{gamma}
\Gamma =& 4 \varepsilon_0^2 \gamma \int_{\eta_p^{\rm min}}^{\eta_p^{\rm max}}
{\rm d}\eta_p \ \eta_p^3 {\eta_h}^2  
 \\
&\times
\sum_{l_p} \left[ l_p \varrho_{l_p-1}(\eta_h) + (l_p+1) \varrho_{l_p+1}(\eta_h) \right] \varrho_{l_p}(\eta_p)  ,
\nonumber
\end{align}
where $\gamma = 2 \pi \hbar^3/3N m_{\rm e}^2 \omega_{\rm M} a^4$, 
$\eta_p^{\rm min} = \eta_{\rm F} \max{(1, \sqrt{\xi})}$, $\eta_p^{\rm max} = \eta_{\rm F} \sqrt{1+\xi}$, 
$\xi = \hbar \omega_{\rm M}/\varepsilon_{\rm F}$ and $\eta_{\rm F} = k_{\rm F}a$.

The SP linewidth depends on the $l$-fixed DOS of the particles and holes. The asymptotic distributions 
of the zeros of the Bessel functions were used in Refs.~\onlinecite{barma} and \onlinecite{yannouleas} to 
obtain the leading behavior of $\Gamma$ for the largest radii of the considered interval. Corrections,
relevant for smaller radii, necessitate numerical or semiclassical approaches.

%=============================================================================
%=============================================================================
%=============================================================================
\subsection{Semiclassical approach and smooth size-dependent component of the plasmon 
linewidth}
The semiclassical approximation to the radial problem (see Appendix \ref{sec:semiclassics}) allows us to write
the $l$-fixed DOS as
\begin{equation}
\label{DOS}
\varrho_l(\varepsilon)  = \frac{\tau_l(\varepsilon)}{2 \pi \hbar} \left\{ 1 + 
 2 \sum_{\tilde r=1}^{\infty} \cos{\left[\tilde r \left( \frac{S_l(\varepsilon)}{\hbar}
 -\frac{3 \pi}{2} \right) \right]} \right\} .
\end{equation}
The classical action of the periodic orbit at energy $\varepsilon$ is 
\begin{align*}
S_l(\varepsilon) = 2 \hbar &\bigg[ \sqrt{{(ka)}^2-{(l+1/2)}^2} 
\nonumber \\
&-\left(l+\frac 12\right) \arccos{\left( \frac{ l +1/2}{ ka}\right)} \bigg] ,
\end{align*}
while its period is given by
\begin{equation*}
\tau_l(\varepsilon) = \frac{\hbar \sqrt{{(ka)}^2-{(l+1/2)}^2}}{\varepsilon_0 {(ka)}^2} ,
\end{equation*}
and we note $\tilde r$ the number of repetitions of the periodic orbit. Within the semiclassical
approximation, the finite height $V_0$ of the self-consistent potential is irrelevant since the classical 
trajectories at a given energy are not sensitive to the shape of the potential above this energy.

In the semiclassical approach of Ref.~\onlinecite{molina}, that we extend and improve in the sequel, it is 
natural to decompose the DOS into a smooth part $\varrho_l^0$ and an oscillating part
$\varrho_l^{\rm osc}$ [Eqs.~\eqref{DOS} and \eqref{trace}]. With Eq.~\eqref{gamma}, this leads
to the dominant (smooth $1/a$-dependent) component of $\Gamma$ (due to the terms 
$\varrho_{l_p}^0 \varrho_{l_h}^0$ of the product) with nonmonotonous (in $a$) corrections.

For the smooth part, we assume that $l_p\gg1$, consistent with the fact that we are interested in leading 
order contributions in $\hbar$. Then we use $l_h \pm 1 \simeq l_p$ and ap\-pro\-xi\-ma\-te the sum over 
$l_p$ by an integral. Setting $y = l_p^2/\eta_{\rm F}^2$ and $z = \eta_p^2/\eta_{\rm F}^2$, we find
\begin{equation}
\label{Gamma^0}
\Gamma^0(a) = \frac{\gamma (k_{\rm F}a)^6}{2 \pi^2} 
\int_{\max{(1, \xi)}}^{1+\xi} \!\!\!\!\!{\rm d}z 
\int_0^{z-\xi} {\rm d}y \sqrt{z-y} \sqrt{z-y-\xi}  .
\end{equation}
Performing the integrals of Eq.~\eqref{Gamma^0} leads to the smooth component $\Gamma^0$ given by 
Eq.~\eqref{sp_smooth}. The $1/a$ dependence agrees with the linear-response result of Kawabata and 
Kubo. \cite{kubo} The function $g$ appearing in Eq.~\eqref{sp_smooth} decreases with $\xi$ with 
$g(0) = 1$ and $\lim_{\xi \rightarrow \infty} g(\xi) = 0$. Its explicit form can be found in 
Refs.~\onlinecite{barma} and \onlinecite{yannouleas}; it is reproduced in Fig.~\ref{functions} of
Appendix \ref{sec_gh}.

\begin{figure}[t]
\begin{center}
\includegraphics[width=8truecm]{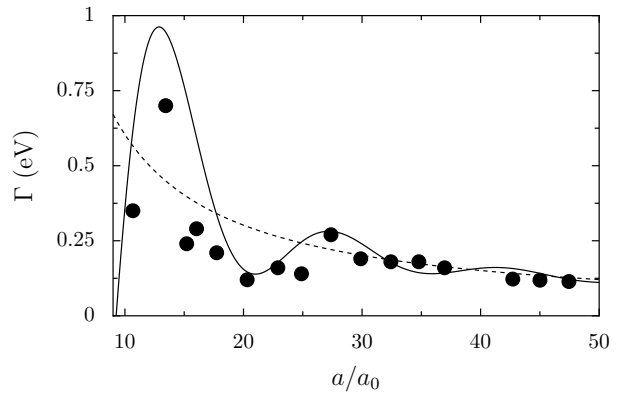}
\caption{\label{fig_gamma} Inverse lifetime of the first collective excitation in Na nanoparticles as a 
function of the radius $a$ of the particle. The dashed line is the smooth  
part of the single plasmon linewidth, Eq.~\eqref{sp_smooth}. The full line is the smooth part plus the
oscillating contribution from Eq.~\eqref{sp_oscillating} for a number of repetitions $\tilde r = 1$. This 
semiclassical result is compared to numerical TDLDA calculations (dots) for clusters with magic numbers 
of atoms between $20$ and $1760$.}
\end{center}
\end{figure}

The smooth component of the linewidth of the collective state is inversely proportional to the radius of the nanoparticle:
This has been interpreted \cite{kubo} as a surface effect arising from the confinement of the
single-particle states. The a\-na\-ly\-ti\-cal evaluation of $\Gamma^0$ agrees with the numerical
calculations (see dashed line of Fig.~\ref{fig_gamma}). Ex\-pe\-ri\-ments on charged alkaline clusters with a
diameter in the range 1--5 nm in vacuum \cite{brechignac} yield a linewidth of the order of $\Gamma \sim 1$ eV.
Although the charged character of those clusters limits the applicability of our model, we note that our calculated value is
smaller, but of the same order of magnitude than the experimental one. This difference might be explained by additional
contributions to the linewidth present in the experiment.

%=============================================================================
%=============================================================================
%=============================================================================
% SIZE-DEPENDENCE OF THE PLASMON LIFETIME: OSCILLATING PART
\subsection{Shell effects and nonmonotonic behavior of the plasmon linewidth}
\label{sec_sp_osc}
The oscillating part of the DOS \eqref{DOS} gives rise to terms of the type
$\varrho_l^0\varrho_{l'}^{\rm osc}$ as well as $\varrho_l^{\rm osc} \varrho_{l'}^{\rm osc}$. The former 
become negligible (in the semiclassical limit of small $\hbar$) in Eq.~\eqref{gamma} because one
integrates a smooth function multiplied by a highly oscillating one. The latter yield 
\begin{align*}
\Gamma^{\rm osc} =&
\frac{4 \gamma}{\pi^2} \int_{\eta_p^{\rm min}}^{\eta_p^{\rm max}} \!\!\!\!\!{\rm d}\eta_p
\eta_p \sum_{\substack{l_p\\ l_h=l_p\pm1}} f_{l_h} 
\prod_{\alpha = p,h} \sqrt{\eta_\alpha^2-(l_\alpha+1/2)^2}  
\nonumber \\
&\times
\sum_{\tilde r_\alpha \geqslant 1} 
\cos{\left[\tilde r_\alpha \left( \frac{S_{l_\alpha}(\eta_\alpha)}{\hbar}-\frac{3\pi}{2} \right) \right]}  ,
\end{align*}
where $f_{l_h} = l_p$ for $l_h=l_p-1$ and $f_{l_h} = l_p+1$ for $l_h = l_p+1$. We can expand the product 
of the two cosines and keep only the contribution in leading order in $\hbar$, neglecting the highly 
oscillating term as a function of the particle and hole actions. We now write this contribution with the 
help of the Poisson summation rule to obtain
\begin{align}
\label{dp_intermediate}
&\Gamma^{\rm osc}  \simeq
\frac{\gamma}{\pi^2} \int_{\eta_p^{\rm min}}^{\eta_p^{\rm max}} {\rm d}\eta_p \eta_p
\sum_{\tilde m = -\infty}^{+\infty} \sum_{\substack{\tilde r_p, \tilde r_h \geqslant 1 \\ \sigma = \pm}}
\int_{-1/2}^{l_{\rm max}} {\rm d}l_p 
\nonumber \\
&\hspace{-.3truecm}\times
\sum_{l_h = l_p \pm 1} f_{l_h}
\prod_{\alpha = p,h} \sqrt{\eta_\alpha^2-(l_\alpha+1/2)^2}
\ {\rm e}^{\sigma {\rm i} \phi_{l_p}^{\tilde r_p \tilde r_h \tilde m}(\eta_p)}  ,
\end{align}
where we have defined the phase 
\begin{align}
\label{phase}
\phi_{l_p}^{\tilde r_p \tilde r_h \tilde m}(\eta_p) = &
\frac{\tilde r_p S_{l_p}(\eta_p)}{\hbar}-\frac{\tilde r_h S_{l_h}(\eta_h)}{\hbar} 
\nonumber \\
&-\frac{3 \pi}{2}(\tilde r_p-\tilde r_h) + 
2\pi \tilde m l_p  .
\end{align}
Performing a stationary phase approximation, given by the condition 
$\left. \partial \phi/\partial l_p \right|_{\bar l_p} = 0$ with the stationary points $\bar l_p$, we obtain the 
stationary phase equation
\begin{equation*}
\tilde r_p \arccos{\left( \frac{\bar{l}_p+1/2}{\eta_p} \right)} 
- \tilde r_h \arccos{\left( \frac{\bar{l}_h+1/2}{\eta_h} \right)} = \pi \tilde m  .
\end{equation*}
The phase of Eq.~\eqref{phase} indicates that the major contribution to the integral over $l_p$ in 
Eq.~\eqref{dp_intermediate} will be given by $\tilde r_p = \tilde r_h$ and $\tilde m = 0$. We then select 
only one point within the full mesh of the stationary points,
\begin{equation}
\label{stationarypoint}
\frac{{\bar l}_p+1/2}{\eta_p} = \frac{\bar{l}_h+1/2}{\eta_h}  .
\end{equation} 
Noticing that $\eta_h = (\eta_p^2-\eta_{\rm F}^2 \xi)^{1/2} < \eta_p$, we see that in order to satisfy 
Eq.~\eqref{stationarypoint}, we have to set ${\bar l}_h = {\bar l}_p-1$. The stationary point is then 
given by
\begin{equation*}
{\bar l}_p = \frac{\eta_p+\eta_h}{2 (\eta_p-\eta_h)}  .
\end{equation*}
Performing the integral over $l_p$ with the help of the stationary phase approximation finally provides 
the following result for the oscillating part of the first plasmon linewidth:
\begin{align}
\label{sp_oscillating}
\Gamma^{\rm osc}(a) =&
6 \sqrt{\pi} \frac{\varepsilon_{\rm F}}{\xi  (k_{\rm F} a)^5}
\int_{\max{(1, \sqrt{\xi})}}^{\sqrt{1+\xi}} {\rm d}\beta \frac{\beta+\beta'}{(\beta-\beta')^4} \nonumber \\
& \times \beta^{5/2} {\beta'}^{3/2} \left[ (k_{\rm F}a)^2 (\beta-\beta')^2-1 \right]^{5/4} \nonumber \\
&\times \sum_{\tilde r = 1}^{\infty} \frac{1}{\sqrt{\tilde r}}
\cos \Bigg\{ 2\tilde r \Big[ \sqrt{(k_{\rm F}a)^2 (\beta-\beta')^2-1} \nonumber \\
& -\arccos{\left( \frac{1}{k_{\rm F}a (\beta-\beta')} \right)} \Big]  - \frac \pi 4 \Bigg\}  ,
\end{align}
where $\beta' = \sqrt{\beta^2-\xi}$ and $\xi = \hbar \omega_{\rm M}/\varepsilon_{\rm F}$. The remaining integral
over $\beta$ can be performed numerically (solid line, Fig.~\ref{fig_gamma}). Assuming that
$k_{\rm F}a \gg 1$, using that $\beta -\beta' \sim 1$ and that the sum over the number of repetitions is 
dominated by the first term, we see that the argument of the cosine is close to $2 k_{\rm F}a$ and 
\begin{equation*}
\Gamma^{\rm osc}(a) \sim \frac{\varepsilon_{\rm F}}{(k_{\rm F} a)^{5/2}} \cos{(2 k_{\rm F} a)}  .
\end{equation*}

Therefore the linewidth of the single SP excitation has a nonmonotonic behavior as a function of the
size $a$ of the metallic cluster. This is due to the density-density particle-hole correlation appearing in the
Fermi Golden Rule \eqref{Gamma_sp}. Let us mention that the result of Eq.~\eqref{sp_oscillating} is 
slightly different from the  one of our previous work. \cite{molina} This is due to the fact that we have 
used here a more rigorous treatment of the semiclassical radial problem. As in Ref.~\onlinecite{molina}, 
we have to set a phase shift in our analytical prediction  of Eq.~\eqref{sp_oscillating} to map the TDLDA
numerical points in Fig.~\ref{fig_gamma}. This is due to the fact that we have taken only one stationary
point [Eq.~\eqref{stationarypoint}] and neglected all the other contributions coming from the full mesh of
stationary points which influence the phase appearing in Eq.~\eqref{dp_intermediate}.

A nonmonotonic behavior has also been observed experimentally in the case of charged lithium
clusters. \cite{brechignac} Our numerical TDLDA calculations (that we have extended here to larger sizes,
up to $1760$ atoms) confirm the presence of size-dependent oscillations for alkaline metals. The 
semiclassical approach also predicts a nonmonotonous behavior of $\Gamma$ for noble metal clusters,
in agreement with recent experimental results. \cite{rodriguez} However, the presence of different 
dielectric constants inside and outside the cluster render the problem more involved. This issue is
discussed in the following section.

%=============================================================================
%=============================================================================
%=============================================================================
%=============================================================================
\section{Plasmon linewidth with an inhomogeneous dielectric environment}
\label{sec_inhomogeneous}
In a previous analysis of the surface plasmon (SP) linewidth, \cite{molina} we were interested by the case of
noble metallic nanoparticles (where the d electrons are modeled via a dielectric constant
$\epsilon_{\rm d}$) embedded in a matrix of dielectric constant $\epsilon_{\rm m}$. The two dielectric 
constants affect $\omega_{\rm M}$ as discussed in the introduction. However, a generalization of the 
derivation of Sec.~\ref{sec_sp} shows that, as long as we work with the hypothesis of a steep potential 
[Eq.~\eqref{staircase}], the smooth part of $\Gamma$ is still given by Eq.~\eqref{sp_smooth}. For a silver
nanoparticle ($\epsilon_{\rm d} \simeq 3.7$) embedded in an argon matrix ($\epsilon_{\rm m} \simeq 1.7$),  
\cite{abeles} using Eq.~\eqref{sp_smooth} yields a value of $\Gamma^0$ about three times larger than the
TDLDA calculations \cite{molina} (themselves in good agreement with existing experiments
\cite{charle}). This discrepancy makes the more systematic study of the dependence of the plasmon 
lifetime on $\epsilon_{\rm d}$ and $\epsilon_{\rm m}$ presented in this section necessary.

\begin{figure}[t]
\begin{center}
\includegraphics[width=8truecm]{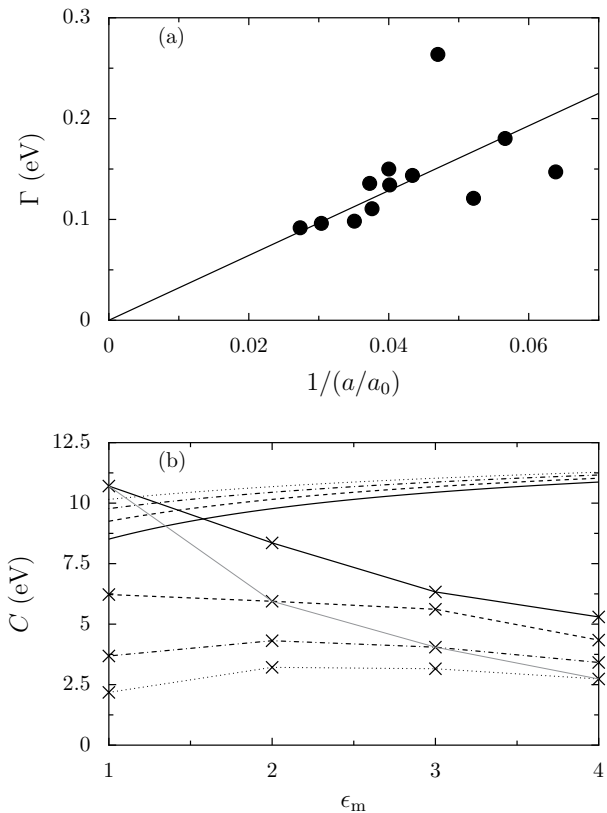}
\caption{\label{fig_phase_diag_Ag}
(a) Surface plasmon linewidth from the TDLDA as a function of the inverse radius for the
example of $\epsilon_{\rm d} = 4$ and $\epsilon_{\rm m} = 2$ (dots). The straight line is a linear fit 
$\Gamma=C/(a/a_0)$.
(b) Prefactor $C$ of the smooth $1/a$ size-dependent component of the surface plasmon linewidth 
$\Gamma^0$ as a function of $\epsilon_{\rm m}$ for  $\epsilon_{\rm d}=1$ (solid line), 
$\epsilon_{\rm d}=2$ (dashed line), $\epsilon_{\rm d}=3$ (dashed-dotted line), and $\epsilon_{\rm d}=4$
(dotted line). The crosses connected by straight lines (guide-to-the-eye) represent the TDLDA 
calculations, while the increasing curves in the upper part of the figure depict the analytical expression 
\eqref{sp_smooth}. The thin gray line is for $\epsilon_{\rm d}=\epsilon_{\rm m}$. The results presented in
the figure are for the electron density of silver ($r_{\rm s} =  3.03$ $a_0$). }
\end{center}
\end{figure}

In Fig.~\ref{fig_phase_diag_Ag}(a) we present the SP linewidth obtained from TDLDA for several particle
sizes between $N=138$ and $1760$, taking $\epsilon_{\rm d} = 4$ and $\epsilon_{\rm m} = 2$ and 
the electron density of silver ($r_{\rm s} = 3.03$ $a_0$). As in the case of Fig.~\ref{fig_gamma}, we see
that for relatively large radii the linewidth can be approximated by $\Gamma^0=C/(a/a_0)$ while for 
smaller radii $a$, superimposed oscillations become noticeable. As shown in  
Fig.~\ref{fig_phase_diag_Ag}(b), when plotting the coefficients $C$ as a function of $\epsilon_{\rm d}$
and $\epsilon_{\rm m}$, we see that the numerical results are at odds with the simple prediction of
Eq.~\eqref{sp_smooth} (upward continuous curves).

The increase of $\Gamma^0$ with $\epsilon_{\rm d}$ and $\epsilon_{\rm m}$ in the latter case 
arises from the fact that the function $g$ is decreasing with 
$\xi = \hbar \omega_{\rm M}/\varepsilon_{\rm F}$ and the Mie frequency 
$\omega_{\rm M} = \omega_{\rm p}/\sqrt{\epsilon_{\rm d}+2\epsilon_{\rm m}}$ is redshifted when
$\epsilon_{\rm d}$ or $\epsilon_{\rm m}$ is increasing. Calculations performed for the electronic density 
of sodium ($r_{\rm s} =  3.93$ $a_0$) give the same kind of discrepancy between Eq.~\eqref{sp_smooth} 
and TDLDA results. \cite{footnote_Na}

The discrepancy between the numerics and Eq.~\eqref{sp_smooth} shows that a direct application of the  
analytical approach presented in the preceeding section does not reproduce the TDLDA results. As we will 
see in the following, the discrepancy is caused by approximating the electronic self-consistent 
potential by a square well.

The TDLDA calculations show that the shape of the self-consistent potential is modified when one 
increases the dielectric constants $\epsilon_{\rm d}$ or $\epsilon_{\rm m}$. In Fig.~\ref{fig_slope}
we present the self-consistent potential of a nanoparticle consisting of $N=832$ atoms 
($r_{\rm s} = 3.03$ $a_0$) for various values of $\epsilon = \epsilon_{\rm d} = \epsilon_{\rm m}$. This
choice does not correspond to a physical realization, but it is useful for the interpretation of the analytical
work, as it merely represents a renormalization of the electronic charge. The main effect of increasing 
$\epsilon$ is the decrease of the slope of the potential near the boundary $r=a$. This indicates that our 
approximation of a square-well potential becomes less valid as the dielectric constant is increased. The 
$\Gamma^0$ dependence on $\epsilon$ in this case is obtained by moving along the line 
$\epsilon_{\rm d} = \epsilon_{\rm m}$ in Fig.~\ref{fig_phase_diag_Ag}(b).

In the following we refine the calculation of the dipole matrix element \eqref{slope_dipole} in order to take 
into account the behavior of the slope of the self-consistent potential. The finite value of the slope of
the self-consistent potential is often ignored. But here, it is necessary to go beyond the hypothesis of 
infinitely steep potential walls [Eq.~\eqref{staircase}] in order to make progress. As it can be seen from
Eq.~\eqref{slope_dipole}, the dipole matrix element is proportional to the matrix element of the derivative 
of the potential $V$ with respect to $r$. In the sequel, we show that below a certain value, the dipole 
matrix element is directly proportional to the slope of the self-consistent potential near the interface and 
estimate the slope from a simple model. Since the linewidth is proportional to the square of the dipole 
matrix element, we see that $\Gamma$ decreases with the slope, and thus with the increase of the 
dielectric constant.

%=============================================================================
%=============================================================================
%=============================================================================
\subsection{Surface plasmon linewidth with a soft self-consistent potential}
In order to improve our understanding of the role of a dielectric mismatch to the SP linewidth, we now 
need to come back to the evaluation of Eq.~\eqref{Gamma_sp} without making the approximation of an 
infinitely steep well for the self-consistent potential. A simplified way of taking into account the 
noninfinite slope
of $V(r)$ is to change Eq.~\eqref{staircase} by
\begin{equation*}
V(r) = 
\begin{cases}
	\displaystyle
	0 ,&\displaystyle  0 \leqslant r < a-\frac{d_s}{2}  , \\
	\displaystyle
	s\left( r-a- \frac{d_s}{2} \right) +V_0  , &\displaystyle
	 a- \frac{d_s}{2} \leqslant r \leqslant a+\frac{d_s}{2}  , \\
	\displaystyle
	V_0  ,&\displaystyle r > a+\frac{d_s}{2}  ,
\end{cases}
\end{equation*}
where the distance $d_s$ on which the slope $s = V_0/d_s$ is nonvanishing is assumed to be small as
compared to the nanoparticle radius $a$. We first need an ap\-pro\-xi\-ma\-tion for the dipole matrix 
element between particle and hole states in that potential. As explained in Appendix \ref{sec:dipol_ME}, this
can be done semiclassically using the limit in which particle and hole states are close in energy 
[$(\varepsilon_p-\varepsilon_h)/\varepsilon_{\rm F} = \hbar \omega_{\rm M}/\varepsilon_{\rm F} \ll 1$].
This semiclassical approximation relates the dipole matrix element to the Fourier components of the 
classical trajectory in the one-dimensional effective potential $V_{l}^{\rm eff}(r)$. As a simplifying 
approximation, we neglect the centrifugal part of the effective potential above $r > a-d_s/2$. Integrating
the classical equation of motion, we obtain periodic trajectories (for $\varepsilon < V_0$) given by
\begin{equation*}
r(t) = 
\begin{cases}
	\displaystyle
	\sqrt{\frac{2 \varepsilon}{m_{\rm e}}t^2+\frac{\hbar^2(l+1/2)^2}{2 m_{\rm e} \varepsilon}}  ,
	&  t \leqslant t_{\rm c}  ,\\
	\displaystyle
	-\frac{s}{2 m_{\rm e}} \left( \frac{\tau_l}{2}-t \right)^2+a+\frac{d_s}{2}-\frac{V_0-\varepsilon}{s}  ,
	& t > t_{\rm c}   ,
\end{cases}
\end{equation*}
with $r(t_{\rm c}) = a-d_s/2$ and where $\tau_l$ is the period. We can now evaluate the dipole matrix 
element using the semiclassical Eq.~\eqref{dipol_ME_sc}, neglecting the acceleration of the particle for  
$r-a+d_s/2 \rightarrow 0^-$ (justified for $a \gg d_s$). An expansion in $1/\Delta n$ (where $\Delta n$ is 
the difference between the radial quantum number of the particle and of the hole) gives, up to an irrelevant 
phase factor
\begin{equation}
\label{ME_slope}
{\cal R}_{k_p k_h}^{l_p l_h} \simeq 
\frac{s}{m_{\rm e}} \frac{2}{\tau_{l_p}} \frac{\hbar^3}{(\varepsilon_p-\varepsilon_h)^3}
\sin{\left( \pi \Delta n \frac{\delta t}{t_{\rm c}} \right)}  ,
\end{equation}
with $\delta t = \tau_{l_p}/2-t_{\rm c}$ the time spent by the particle in the region where the slope is
nonvanishing.

An estimation of the argument of the sine gives $(\hbar \omega_{\rm M}/\Delta) (d_s/a)$, with $\Delta$ 
the mean level spacing. Typical values give $\hbar \omega_{\rm M}/\Delta \sim 10^4 \gg 1$. In the limit
of a very large slope, $d_s/a$ tends to zero. Then, the argument of the sine is very small compared to one, 
and we recover the semiclassical evaluation of Eq.~\eqref{dipol_ME_sc_infiniteslope} with an infinite 
slope. On the contrary, if we assume that $d_s$ is of the order of the spillout length \cite{madjet} 
($\sim a_0$), the argument of the sine is much greater than one. Inserting Eq.~\eqref{ME_slope} into
Eq.~\eqref{Gamma_sp}, we obtain
\begin{equation*}
\Gamma^0 = \frac{2 s^2}{\pi \hbar \omega_{\rm M}^3 N m_{\rm e}} \!\!\!\!
\int \limits_{\varepsilon_{\rm F}}^{\varepsilon_{\rm F}+\hbar \omega_{\rm M}} \!\!\!\!
{\rm d} \varepsilon_{p} \sum_{\substack{l_p, m_p\\l_h, m_h}} \left( {\cal A}_{l_pl_h}^{m_pm_h} \right)^2
\sin^2{\left( \pi \Delta n \frac{\delta t}{t_{\rm c}} \right)}  .
\end{equation*}
Averaging the highly oscillating sine (squared) by $1/2$ gives for the SP linewidth 
in the limit $\xi \rightarrow 0$
\begin{equation}
\label{Gamma_slope}
\Gamma^0(a) \simeq \frac{3 s^2}{4} \frac{1}{m_{\rm e} \omega_{\rm M}^2} \frac{1}{k_{\rm  F}a}  .
\end{equation}
We then see that in the case of a soft self-consistent potential, the SP linewidth is proportional to the 
square of the slope $s$ of that potential. When one increases the dielectric constant of the 
medium, the slope decreases (see Fig.~\ref{fig_slope}) and therefore $\Gamma^0$ decreases. We also
notice that the smooth $1/a$ size dependence of the SP linewidth remains valid even for a finite slope.

%=============================================================================
%=============================================================================
%=============================================================================
\subsection{Steepness of the self-consistent potential with a dielectric mismatch}
In order to estimate the slope of the self-consistent potential, we consider the simpler geometry of a
metallic slab of dielectric constant $\epsilon_{\rm d}$, bounded by two interfaces at $x = \pm w/2$ and 
with an infinite extension in the $(y,z)$ plane, surrounded by a dielectric medium with a constant 
$\epsilon_{\rm m}$. This geometry allows us to simplify the problem to an effective one-dimensional 
system and can be expected to provide a good approximation for the shape of the potential near the 
interface for the sphere geometry. 

We make the jellium approximation for the ionic density $n_{\rm i} (x) = n_{\rm i} \Theta(w/2-|x|)$, with
$\Theta$ the step function, and work within the Thomas-Fermi approach, writing the local energy in the electrostatic field $\phi$ as
\begin{equation*}
\varepsilon = \frac{p^2(x)}{2 m_{\rm e}}-e \phi(x)  ,
\end{equation*}
and the electronic density (at zero temperature) as 
\begin{equation*}
n_{\rm e}(x) = 
\frac{1}{3 \pi^2} \left( \frac{2 m_{\rm e}}{\hbar^2} \right)^{3/2}
\left[ \mu+e \phi(x)\right]^{3/2}  ,
\end{equation*}
with $\mu$ the chemical potential in the potential $V(x) = -e \phi(x)$. The Thomas-Fermi approach to 
surfaces is known to have serious shortcomings \cite{desjonqueres} (for instance, it predicts a vanishing 
work function). However, it will be useful for our estimation of the slope of the mean field seen by the 
charge carriers. The self-consistency is achieved through the Poisson equation
\begin{equation}
\label{poisson}
\frac{{\rm d}^2 \phi}{{\rm d}x^2} =
\begin{cases}
	\displaystyle
	\frac{4 \pi e}{\epsilon_{\rm d}} \left[ n_{\rm e}(x)-n_{\rm i} \right] ,
	& \displaystyle |x| < \frac w2  , \\
	\displaystyle
	\frac{4 \pi e}{\epsilon_{\rm m}}  n_{\rm e}(x)  ,
	& \displaystyle   |x| > \frac w2  .
\end{cases}
\end{equation}

First we consider the simpler case where $\epsilon_{\rm d} = \epsilon_{\rm m}=\epsilon$. In this case, integrating
once Eq.~\eqref{poisson} and invoking the continuity of the potential and the electrical field,
we find for the slope of the self-consistent field at $x=w/2$
\begin{equation}
\label{slope_e}
s = \frac{4 e}{\sqrt{15 \pi}} \left( \frac{2 m_{\rm e}}{\hbar^2} \right)^{3/4} \frac{\mu_1^{5/4}}{\epsilon^{7/4}} 
\left[ 1- \frac{2}{5 \epsilon^{3/2}} \left( \frac{\mu_1}{\varepsilon_{\rm F}} \right)^{3/2} \right]^{5/4}  ,
\end{equation} 
where we have assumed the scaling $\mu \approx \mu_1/\epsilon$ with $\mu_1$ the chemical potential 
in the case where $\epsilon = 1$, and $\varepsilon_{\rm F}$ is the Fermi energy of the free electron gas. 
The chemical potential is fixed by the consistency condition
\begin{equation}
\label{fermi_level}
\sqrt{\frac{\epsilon \mu}{8 \pi e^2 n_{\rm i}}} 
\int_{1-\frac 25 \left( \frac{\mu}{\varepsilon_{\rm F}} \right)^{3/2}}^1
\frac{{\rm d} u}{f(u)} = \frac w2 
\end{equation}
with 
\begin{equation*}
f(u) = \sqrt{\frac 25 \left( \frac{\mu}{\varepsilon_{\rm F}} \right)^{3/2} (u^{5/2}-1)-(u-1)}  .
\end{equation*}
If we do not have any dielectric constant (i.e., $\epsilon = 1$), the same equation is obtained but without
the prefactor $\sqrt{\epsilon}$. The integral in Eq.~\eqref{fermi_level} is clearly dominated by its prefactor. 
Then, assuming that the integral appearing in this equation does not change appreciably when we have a 
dielectric constant, we find the scaling $\mu \approx \mu_1/\epsilon$. Therefore, we see that the slope at 
the interface is decreasing with increasing values of the dielectric constant $\epsilon$, a feature 
confirmed by our TDLDA calculations (see Fig.~\ref{fig_slope}).

In the case where we have a dielectric mismatch between the metallic slab and the environment, the 
continuity of the normal component of the displacement field $\bf D$ gives, perturbatively, in the limit  
$|\epsilon_{\rm d}-\epsilon_{\rm m}| \rightarrow 0$, 
\begin{align}
\label{slope_edem}
s =&
\frac{4 e}{\sqrt{15 \pi}} \left( \frac{2 m_{\rm e}}{\hbar^2} \right)^{3/4}
\frac{\mu_1^{5/4}}{\epsilon_{\rm m}^{1/2} \epsilon_{\rm d}^{5/4}}
\left[ 1- \frac{2}{5 \epsilon_{\rm d}^{3/2}}\left( \frac{\mu_1}{\varepsilon_{\rm F}} \right)^{3/2} \right]^{5/4} 
\nonumber \\
& \times
\left\{1+\frac{\epsilon_{\rm d}-\epsilon_{\rm m}}{2 \epsilon_{\rm d}^{5/2}} 
\left(\frac{\mu_1}{\varepsilon_{\rm F}} \right)^{3/2}
\left[ 1-\frac{2}{5 \epsilon_{\rm d}^{3/2}} \left( \frac{\mu_1}{\varepsilon_{\rm F}} \right)^{3/2} \right] \right\}  ,
\end{align}
with the scaling $\mu \approx \mu_1/\epsilon_{\rm d}$, which can be justified in the same 
manner as for the case of a single dielectric constant. The only difference is that in the case of a dielectric 
mismatch, we obtain Eq.~\eqref{fermi_level}, up to a change of $\epsilon$ by $\epsilon_{\rm d}$. We then 
see that the slope $s$ of the confining mean-field potential at the interface is decreasing either with 
$\epsilon_{\rm d}$ or $\epsilon_{\rm m}$ (for small $|\epsilon_{\rm d}-\epsilon_{\rm m}|$), in agreement 
with the TDLDA calculations. 

This Thomas-Fermi approach to the mean-field potential of a metallic slab then provides 
an estimate of the slope of that potential near the interface between the slab and the surrounding 
environment. It can be expected that these results are also applicable to the more involved problem of the 
metallic sphere, up to some geometrical prefactors. In the following section, we will incorporate our 
estimate of the self-consistent potential slope in our evaluation of the SP lifetime.

%=============================================================================
%=============================================================================
%=============================================================================
\subsection{Surface plasmon linewidth with a dielectric mismatch}
We can now use our estimate \eqref{slope_edem} for the slope of the self-consistent potential in our 
evaluation \eqref{Gamma_slope} of the SP linewidth. In order to do that, we assume that the chemical potential $\mu_1$
for $\epsilon = 1$ is the Fermi energy $\varepsilon_{\rm F}$ of a free electron gas.

In the case where we have a charge renormalization (i.e., $\epsilon_{\rm d} = \epsilon_{\rm m} = \epsilon$),
we obtain by inserting Eq.~\eqref{slope_e} into Eq.~\eqref{Gamma_slope}  
\begin{equation}
\label{Gamma_e}
\Gamma^0(a) \simeq \frac{9}{5} \frac{\varepsilon_{\rm F}}{k_{\rm F} a} \frac{1}{\epsilon^{5/2}} 
\left( 1-\frac{2}{5 \epsilon^{3/2}} \right)^{5/2}  .
\end{equation}
This result qualitatively reproduces the decrease obtained from TDLDA for $\Gamma^0\, a/a_0$ as a 
function of the dielectric constant $\epsilon$ as it can be seen on Fig.~\ref{fig_phase_diag_Ag}(b) (gray
thin line). We notice that for $\epsilon = 1$, we have $\Gamma^0 \approx \varepsilon_{\rm F}/k_{\rm F} a$ 
in the limit of small $\xi$, which has to be compared with Eq.~\eqref{sp_smooth} giving 
$1.5 \ \varepsilon_{\rm F}/k_{\rm F} a$. This small discrepancy is not surprising, regarding the various 
approximations we made here. 

In the case where we have a dielectric mismatch, by inserting Eq.~\eqref{slope_edem} into
Eq.~\eqref{Gamma_slope} and making the expansion for small
$\Delta \epsilon = \epsilon_{\rm d}-\epsilon_{\rm m}$,  we obtain
\begin{equation}
\Gamma^0(a) \simeq \Gamma_{(\Delta \epsilon = 0)}^0(a) + A \Delta \epsilon
\end{equation}
for fixed $\epsilon_{\rm d}$ 
and 
\begin{equation}
\label{Gamma_m}
\Gamma^0(a) \simeq \Gamma_{(\Delta \epsilon = 0)}^0(a) - B \Delta \epsilon
\end{equation}
for fixed $\epsilon_{\rm m}$. In the above two equations, $A$ and $B$ are two positive
coefficients not specified here, and $\Gamma_{(\Delta \epsilon = 0)}^0$ is given by Eq.~\eqref{Gamma_e}. 
These results confirm the behavior of the TDLDA calculations depicted on Fig.~\ref{fig_phase_diag_Ag}(b)
around $\Delta \epsilon=0$ (thin gray line).
For instance, if we are at $\epsilon_{\rm m}$ fixed, we see that when $\Delta \epsilon > 0$, 
Eq.~\eqref{Gamma_m} predicts that $\Gamma^0\, a/a_0$ decreases for increasing value of 
$\epsilon_{\rm d}$. 

We have shown in this section how to take into account an inhomogeneous dielectric 
environment in our semiclassical model through the corrections in the slope of the mean-field potential. 
This improved theory is in qualitative agreement with the TDLDA calculations.

%=============================================================================
%=============================================================================
%=============================================================================
%=============================================================================
% HIGHER COLLECTIVE EXCITATIONS: DOUBLE PLASMON
\section{Higher collective excitations: double plasmon}
\label{sec_dp}
In this section we discuss the lifetime of the second collective excitation level in metallic nanoparticles.
Although there is no clear direct experimental observation of a double plasmon in metallic
clusters, the development of femtosecond spectroscopy will certainly allow for detailed studies in the near
future. Recent experiments observed the ionization of the charged cluster Na$_{93}^+$ by a femtosecond 
laser pulse and claimed it was a consequence of the excitation of the second plasmon state. 
\cite{schlipper}$^{\rm a}$ However, the analysis of the distribution of photoelectrons yielded a thermal 
distribution and therefore the relevance of the double plasmon for this experiment is not yet settled. 
\cite{schlipper}$^{\rm b}$ On the other hand, it is clear that a strong-enough laser pulse will excite the 
second collective state. Such an excitation will be a  well-defined resonance only if its linewidth is small 
compared with other scales of the photoabsorption spectrum (like for instance $\omega_{\rm M}$).
 
Second collective excitations have been widely a\-na\-ly\-zed in the context of giant dipolar resonances in
nuclei. \cite{gu} The anharmonicities were found to be relatively small, making it possible to observe 
this resonance. \cite{ritman} The theoretical tools developed in nuclear physics have been adapted 
to the study of the double plasmon in metallic clusters. \cite{catara, hagino} In particular, a variational 
approach \cite{hagino} showed that the difference between the energy of the double plasmon and 
$2 \hbar \omega_{\rm M}$ decreases as $N^{-4/3}$ with the size of the nanoparticle. In our calculations, 
we will assume that the double-plasmon energy is exactly twice the Mie energy.

For most of the clusters of experimental interest, $2 \hbar \omega_{\rm M} > W > \hbar \omega_{\rm M}$, 
where $W$ is the work function. Io\-ni\-za\-tion then appears as an additional decay channel of the 
double plasmon that competes with the Landau damping, while it is not possible if only the single 
plasmon is excited. 
\cite{koskinen}

%=============================================================================
%=============================================================================
%=============================================================================
% SECOND PLASMON DECAY FOR STRONGLY CONFINED ELECTRONS
\subsection{Second plasmon decay: Landau damping}
In this section, we consider processes which do not lead to ionization, that is, the final particle energies 
verify $\varepsilon_p < V_0 = \varepsilon_{\rm F}+W$. A sufficiently strong laser excitation gives rise to an 
initial center-of-mass state which is a linear superposition of the ground-state $| 0_{\rm cm} \rangle$, the 
first ($| 1_{{\rm cm}, Z} \rangle$), and the second ($| 2_{{\rm cm}, Z} \rangle$) harmonic oscillator excited
states. 

The second plasmon state can decay by two distinct Landau damping processes. A first-order process, 
with a rate $\Gamma_{2 \rightarrow 1}$, results from the transition of $| 2_{{\rm cm}, Z} \rangle$ 
(double plasmon) into $| 1_{{\rm cm}, Z} \rangle$ (single plasmon). The corresponding matrix element 
of the perturbation $\hat H_{\rm c}$ between these two states is a factor of $\sqrt{2}$ larger than 
the one worked in Sec.~\ref{sec_PPA}, and then $\Gamma_{2 \rightarrow 1} = 2 \Gamma$ [where
$\Gamma$ is the single-plasmon linewidth given by Eq.~\eqref{Gamma_sp} and calculated under certain 
approximations in Sec.~\ref{sec_sp}]. Thus, the contribution of the first-order process to  the linewidth is
just twice that of the single plasmon, and shows the same nonmonotonic features superposed to a $1/a$
size-dependence. 

The other mechanism one has to take into account is the second-order process, where the 
double plasmon decays directly into the center-of-mass ground state. This is possible provided that
$V_0 > 2 \hbar \omega_{\rm M}$. In order to simplify the calculation we assume, for the remaining of this 
section, that $V_0 \rightarrow \infty$. The corresponding linewidth $\Gamma_{2 \rightarrow 0}$ is given
by the Fermi Golden Rule in second order in perturbation theory by \cite{bertsch:dp}
\begin{widetext}
\begin{equation*}
\Gamma_{2 \rightarrow 0} = 2 \pi \sum_{f_{\rm MF}} 
\left| \sum_{f'_{\rm MF}} 
\frac{\langle 0_{\rm cm}, f_{\rm MF} | \hat H_{\rm c} | 1_{{\rm cm}, Z},  f'_{\rm MF} \rangle 
\langle 1_{{\rm cm}, Z}, f'_{\rm MF} | \hat H_{\rm c} | 2_{{\rm cm}, Z},  0_{\rm MF} \rangle 
 }{\hbar \omega_{\rm M}-\varepsilon_{f'_{\rm MF}}}
 \right|^2 \delta(2 \hbar \omega_{\rm M}-\varepsilon_{f_{\rm MF}}) .
\end{equation*}
\end{widetext}
Expliciting the perturbation \eqref{Hc} and restricting ourselves to the random phase approximation which 
allows only one particle-one hole transitions, we obtain
\begin{equation}
\label{FGR_bertsch}
\Gamma_{2 \rightarrow 0} = \frac{\pi \hbar^2 \omega_{\rm M}^6 m_{\rm e}^2}{N^2} \sum_{ph} \left| K_{ph} \right|^2 \delta(2 \hbar \omega_{\rm M}
-\varepsilon_p+\varepsilon_h)  ,
\end{equation}
with
\begin{equation}
\label{Kph}
K_{ph} =  \sum_{i \neq p,h}
\frac{d_{pi} d_{ih}}{\hbar \omega_{\rm M}- \varepsilon_{i} + \varepsilon_h}  .
\end{equation}
The sum over $i$ runs over all the virtual intermediate states. We use the same notations as in
Sec.~\ref{sec_sp} and replace the sums over particle and hole states by integrals over the energy with the 
appropriate density of states (DOS), which is approximated by its semiclassical counterpart.

As in the case of the single plasmon, we work in the limit $l_p \gg 1$ in order to find the smooth
size-dependent contribution $\Gamma_{2 \rightarrow 0}^0$ (and the corresponding $K_{ph}^0$). Using 
Eqs.~\eqref{dab}, \eqref{sp_angular}, \eqref{sp_radial}, and the selection rules, we have 
\begin{align}
\label{Kbar}
&K_{ph}^0 = \frac{8 \hbar^2}{\pi m_{\rm e} \varepsilon_0^2} 
\left[  
{\cal A}_{l_pl_{p}+1}^{m_pm_p} {\cal A}_{l_{p}+1l_h}^{m_pm_h}  I_{l_p+1}(\eta_p, \eta_h)
(\delta_{l_hl_p}+\delta_{l_h,l_{p}+2}) \right. \nonumber \\
&\left. +{\cal A}_{l_pl_{p}-1}^{m_pm_p} {\cal A}_{l_{p}-1l_h}^{m_pm_h}  I_{l_p-1}(\eta_p, \eta_h)
(\delta_{l_hl_p}+\delta_{l_h,l_{p}-2})
\right] \delta_{m_pm_h}
  .
\end{align}
Here, we have defined the integral 
\begin{align*}
&I_{l_i}(\eta_p, \eta_h) = 
\\
&
\int_{l_i+1/2}^\infty {\rm d}\eta_i \frac{\eta_i \sqrt{\eta_i^2-(l_i+1/2)^2}}
{\left( \xi \eta_{\rm F}^2 -\eta_i^2+\eta_h^2 \right)
\left[ (\eta_i^2-\eta_p^2) (\eta_i^2-{\eta_h}^2) \right]^2}  . \nonumber
\end{align*}
Inserting Eq.~\eqref{Kbar} into Eq.~\eqref{FGR_bertsch}, replacing the sum over $l_p$ by an integral, we
find 
\begin{align*}
\Gamma_{2 \rightarrow 0}^0 =&
2 \varsigma \int_{\eta_{\rm F}\max{(1, \sqrt{2 \xi})}}^{\eta_{\rm F} \sqrt{1+2\xi}}
{\rm d}\eta_p \ \eta_p \\
&\times
\int_0^{\eta_h} {\rm d}l_p \ l_p
\sqrt{\eta_p^2-l_p^2} \sqrt{{\eta_h}^2-l_p^2} \left[ I_{l_p}(\eta_p, \eta_h) \right]^2  \nonumber  ,
\end{align*}
where the factor of $2$ accounts for the spin degeneracy. We have introduced 
$\varsigma = 64 (\hbar \omega_{\rm M})^6/5 \pi^3 N^2\varepsilon_0^5$ and
$\eta_h = (\eta_p^2-2\eta_{\rm F}^2\xi)^{1/2}$. With the change of variables 
$z = \eta_p^2/\eta_{\rm F}^2$, 
$y = l_p^2/\eta_{\rm F}^2$, and $x = \eta_i^2/\eta_{\rm F}^2$, we obtain 
\begin{equation}
\label{2P_academic}
\Gamma_{2 \rightarrow 0}^0(a) \simeq \frac{81}{10 \pi^3} \frac{\varepsilon_{\rm F}}
{\left( k_{\rm F} a \right)^2} h(\xi)  ,
\end{equation} 
where the function $h(\xi)$ of the parameter $\xi = \hbar \omega_{\rm M}/\varepsilon_{\rm F}$ is
smoothly increasing with $h(0) = 0$. An approximate expression of $h$ is given in Appendix \ref{sec_gh}.

The total linewidth of the Landau damped second plasmon is the sum of the first- and second-order
processes: $\Gamma_{\rm DP} = \Gamma_{2 \rightarrow 1}+\Gamma_{2 \rightarrow 0}$. The different (smooth)
size dependence of both processes [$v_{\rm F}/a$ for the former and $(k_{\rm F} a)^{-1} v_{\rm F}/a$ for
the latter] implies that, except for the smallest clusters, the second-order process gives a ne\-gli\-gi\-ble
contribution to the linewidth of the double plasmon (in comparison with that of the first order). We
might ask the question of whether the inclusion of the oscillating components of both linewidths can
affect the above conclusion in this range of particle sizes. An extension of the calculations 
presented in Sec.~\ref{sec_sp_osc} shows \cite{weick} that the oscillating part of the second-order 
channel of the double plasmon is given by
\begin{equation*}
\Gamma_{2 \rightarrow 0}^{\rm osc}(a) \sim 
\frac{\varepsilon_{\rm F}}{(k_{\rm F}a)^{11/2}} \cos{(2 k_{\rm F}a)} .
\end{equation*}
As indicated before, $\Gamma^{\rm osc}_{2 \rightarrow 1}$ is given by twice the result of 
Eq.~\eqref{sp_oscillating}, therefore these nonmonotonic contributions cannot lead to a significant modification
of our conclusion about the irrelevance of the second-order term for the sizes of physical
interest. We also notice that $\Gamma_{\rm DP} \ll \hbar \omega_{\rm M}$, since for typical nanoparticles,
$\varepsilon_{\rm F} \sim \hbar \omega_{\rm M}$ and $k_{\rm F} a \gg 1$. Therefore, the Landau damping 
is not capable of ruling out the second plasmon as a well-defined resonance.

The lifetime of the second plasmon for the Landau damping processes is simply $\hbar \Gamma_{\rm DP}^{-1}$.
From the experimental point of view, what is usually more relevant is the time it takes for the double
excited state of the center-of-mass system to return to its ground state rather than the lifetime of the 
excited state. Therefore, we also have to take into account the decay of the first plasmon into the 
ground-state $\Gamma_{1 \rightarrow 0}$. If we assume that the recombination of particle-hole pairs
(created by the decay of the double plasmon into the single plasmon) is very fast as compared to other 
time scales, we have  $\Gamma_{1 \rightarrow 0} = \Gamma$. Due to the fact that lifetimes are additive,
we have for this sequential decay a lifetime 
$\tau_{2 \rightarrow 1 \rightarrow 0 } = 1.5 \hbar \Gamma^{-1}$.

%=============================================================================
%=============================================================================
%=============================================================================
% IONIZATION PROCESS
\subsection{Second plasmon decay and ionization}
\begin{figure}[h]
\begin{center}
\includegraphics[width=8truecm]{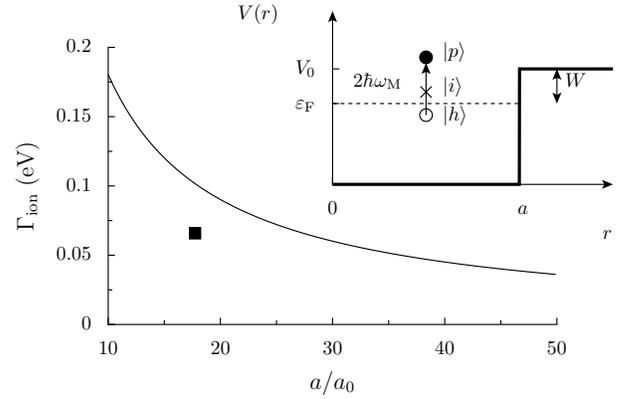}
\caption{\label{fig_gamma_ion} Ionization linewidth of the second plasmon state as a function of the
nanoparticle radius for singly charged Na clusters. Square: experimental value for Na$_{93}^+$ taken from
Ref.~\onlinecite{schlipper}. We have assumed a constant work function
$W = 4.65$ eV and took the experimental Mie frequency of $2.75$ eV.
Inset: scheme of the ionization process of the double-plasmon state which decays by creating
a particle-hole pair of energy $2 \hbar \omega_{\rm M}$, via the intermediate state $| i \rangle$.
Since the energy of the particle is such that $\varepsilon_p > V_0$, ionization occurs.}
\end{center}
\end{figure}

We now examine the last decay channel of the second plasmon state: the relaxation of this collective 
excitation by ejecting an electron from the nanoparticle (ionization, see inset of Fig.~\ref{fig_gamma_ion}).
We now need to determine the particle and hole states in the self-consistent field [Eq.~\eqref{staircase}]
which has a finite height $V_0$, since the ionization process requires the states of the continuum. For
simplicity, we will neglect the Coulomb tail seen for $r > a$ by electrons with an energy 
$\varepsilon_p > V_0$.

In order to determine the particle and hole states, we close the system into a spherical box of radius 
$L \gg a$ to quantize the states above the well and take the limit of $L \rightarrow \infty$ at the end of 
our calculations. We need to do some approximations in order to simplify this difficult problem. First, 
in the high energy limit, we assume that $kr \gg 1$, and then use the asymptotic expansions of the  
quantum mechanical single-particle states inside and outside the well. Even though this approximation 
strongly affects the wave functions near $r = 0$, its impact on the dipole matrix elements is very small. 
\cite{footnote} Second, for the states with energy $\varepsilon < V_0$, we neglect the exponential decay 
of the wave function for $r > a$. Finally, in the spirit of the scattering theory,  we use a simplifed expression
for the normalization of the free states above the well. The above assumptions result in the following 
radial wave functions inside the well ($\varepsilon < V_0$)
\begin{equation*}
u_{kl}^<(r) \simeq 
\begin{cases}
	\displaystyle \sqrt{\frac{2}{a}} \sin{(kr-l \pi/2)}  , & r \leqslant a , \\
	\displaystyle  0  , & r > a  .
\end{cases}
\end{equation*}
The wave-vector $k = (2 m_{\rm e} \varepsilon)^{1/2}/\hbar$ is given by the quantization condition
$ka = l \pi/2+n\pi$, with $n$ a non-negative integer. Outside of the well ($\varepsilon > V_0$), we have
\begin{equation*}
u_{kl}^>(r) \simeq \sqrt{\frac{2}{L}} 
\begin{cases}
	\displaystyle  \alpha_l(k) A(k) \sin{(kr-l \pi/2)}  , & r \leqslant a  , \\
	\displaystyle \sin{\left[ \kappa (r-L) \right]} , & r > a  ,
\end{cases}
\end{equation*}
with $\kappa = (k^2-2 m_{\rm e} V_0/\hbar^2)^{1/2}$. We have introduced the abbreviations
\begin{align*}
	\alpha_l(k) &= {\rm sign} 
	\left\{ \frac{\sin{\left[ \kappa (a-L) \right]}}{\sin{\left( ka-l \pi/2 \right)}} \right\} ,
	\\
	A(k) &= \sqrt{\sin^2{\left[ \kappa (a-L) \right]} 
	+ \left( \frac{\kappa}{k} \right)^2 \cos^2{\left[ \kappa (a-L) \right]}}  .
\end{align*}

The ionization rate of the double-plasmon state $\Gamma_{\rm ion}$ is given by Eq.~\eqref{FGR_bertsch}
in the case where the final particle states $p$ of the sum are in the continuum. Since the effective
(second-order) matrix element $K_{ph}$ [Eq.~\eqref{Kph}] is given by a sum over intermediate states $i$,
we now have contributions from cases where $i$ lies in the well as well as  in the continuum.

When $i$ represents a state in the well, using the angular momentum selection rules, we can write in the 
limit $kr \gg~1$
 \begin{equation}	
\label{ion_ME_1}
{\cal R}_{k_i k_h}^{l_i l_h} =
\frac{(-1)^{n_i-n_h}}{\Delta k_{ih}^2a} \delta_{l_i, l_h \pm 1}   ,
\end{equation}
and 
\begin{align}
\label{ion_ME_2}
{\cal R}_{k_p k_i}^{l_p l_i} =& \pm \sqrt{\frac{a}{L}} \frac{\alpha_{l_p}(k_p) A(k_p)}{ \Delta k_{pi}}
\nonumber \\
&\times
\left[ \cos{\left( \Delta k_{pi}a \right)} - \frac{\sin{\left( \Delta k_{pi}a \right)}}{\Delta k_{pi}a} \right]
\delta_{l_i, l_p \pm 1}  ,
\end{align}
where $\Delta k_{\alpha \beta} = k_\alpha-k_\beta$ ($\alpha, \beta = p, h, i$).

When $i$ represents a state in the continuum, ${\cal R}_{k_i k_h}^{l_i l_h}$ can be obtained by 
exchanging ($p \leftrightarrow i$) and ($i \leftrightarrow h$) in Eq.~\eqref{ion_ME_2}. For the remaining 
case, we have
\begin{align}
\label{ion_ME_4}
&{\cal R}_{k_p k_i}^{l_p l_i} \simeq 
 \frac{a}{L} 
\frac{\alpha_{l_p}(k_p) \alpha_{l_p+1}(k_i) A(k_p) A(k_i)}{\Delta k_{pi}}
\\
&\times
\left[
\cos{\left( \Delta k_{pi}a \right)} -\frac{\sin{\left( \Delta k_{pi}a \right)}}{\Delta k_{pi}a}
\right]
\delta_{l_i, l_p \pm 1}+B(k_p, k_i)  ,
\nonumber
\end{align}
where
\begin{align*}
B(k_p, k_i) &=
\frac{a^2}{L}
\bigg\{
\cos{\left( \Delta \kappa_{pi} L \right)}-\sin{\left( \Delta \kappa_{pi}L \right)} {\rm ci} (|\Delta \kappa_{pi}|a)
\\
&+ \Delta \kappa_{pi}a
\big[
\cos{\left( \Delta \kappa_{pi}L \right)} {\rm sign} (\Delta \kappa_{pi}) {\rm si} (|\Delta \kappa_{pi}|a)
\big]
\bigg\}
 ,
\nonumber
\end{align*}
with si and ci the sine and cosine integral functions.

The semiclassical $l$-fixed smooth DOS can be approximated by
\begin{equation*}
\varrho_l^0(\varepsilon) \simeq \frac{1}{2 \pi \varepsilon_0}
\begin{cases} 
	\displaystyle \frac{\sqrt{(ka)^2-(l+1/2)^2}}{(ka)^2}  ,& \varepsilon < V_0  ,  \\
	\displaystyle  \frac{\sqrt{(\kappa L)^2-(l+1/2)^2}}{(\kappa a)^2}  , & \varepsilon > V_0  .
\end{cases}
\end{equation*}

There is an obvious divergency that occurs in the sum of Eq.~\eqref{Kph} for
$\varepsilon_i = \varepsilon_h$, as it can be seen on the matrix element \eqref{ion_ME_1}. However, a
careful analysis shows that the contribution around that divergency vanishes because of the alternating 
sign when one integrates over $n_i$. For $\varepsilon_i = \varepsilon_p$, there is no divergency in 
Eq.~\eqref{ion_ME_4}. Therefore the dominant contribution to $K_{ph}$ is given by the divergency of the 
term $1/(\hbar \omega_{\rm M}-\varepsilon_i+\varepsilon_h)$ that occurs for $\varepsilon_i < V_0$ in the 
regime we are interested in ($\hbar \omega_{\rm M} < W < 2 \hbar \omega_{\rm M}$). We then have for the 
ionization rate
\begin{align*}
\Gamma_{\rm ion} \simeq& 2  \frac{\pi \hbar^2 \omega_{\rm M}^6 m_{\rm e}^2}{N^2}
\sum_{\substack{p > V_0 \\ h< \varepsilon_{\rm F}}} 
\delta(2 \hbar \omega_{\rm M}-\varepsilon_{p} + \varepsilon_h) \nonumber \\
&\times
\sum_{i, j < V_0} \frac{d_{pi} d_{ih}}{\hbar \omega_{\rm M}- \varepsilon_{i} + \varepsilon_h} 
\frac{d_{pj} d_{jh}}{\hbar \omega_{\rm M}- \varepsilon_{j} + \varepsilon_h} 
 ,
\end{align*}
where the factor of $2$ accounts for the two spin channels and the $d_{\alpha \beta}$ are given by 
Eq.~\eqref{dab} with the approximations \eqref{ion_ME_1} and \eqref{ion_ME_2} for the radial matrix 
elements. Furthermore, we can distinguish in the above equation two contributions: off-diagonal terms 
($i \neq j$) which have divergencies of the principal value type and that we neglect here, and diagonal 
terms ($i=j$) yielding divergencies which determine  $\Gamma_{\rm ion}$. We smooth out the energy 
$\varepsilon_i$ appearing in the denominator by introducing an imaginary part of the order of the mean
level-spacing
\begin{equation*}
\Delta = \frac{3\pi \varepsilon_0^{3/2}}{\sqrt{\hbar \omega_{\rm M} + \varepsilon_h}} 
\end{equation*} 
at an energy $\hbar \omega_{\rm M} + \varepsilon_h$.
This standard procedure of smoothing the divergencies is of critical importance, and that is why in 
Ref.~\onlinecite{bertsch:dp} the final result is presented as a function of $\Delta$. Summing over $l_i$ 
and $m_i$, the remaining sum over the radial quantum numbers $n_i$ can be done with the help of
\begin{equation*}
\sum_{n_i} \frac{1}{\left| \hbar \omega_{\rm M} -\varepsilon_i+\varepsilon_h+{\rm i} \Delta \right|^2} 
\approx \frac{\pi}{4 \Delta} \frac{1}{\hbar \omega_{\rm M}+\varepsilon_h}  .
\end{equation*}
For the smooth terms of the sum, we have taken their values at the divergency to obtain
\begin{align*}
&\Gamma_{\rm ion} \simeq \frac{\pi^2 a (\hbar \omega_{\rm M})^6 \varepsilon_0}{120 N^2 L}
\sum_{l_p} l_p
\int\limits_{\max{(V_0, 2 \hbar \omega_{\rm M})}}^{\varepsilon_{\rm F}+2 \hbar \omega_{\rm M}}
{\rm d} \varepsilon_p \frac{\varrho_{l_p}(\varepsilon_p) \varrho_{l_p}(\varepsilon_h)}{\Delta} \nonumber \\
&\times
\frac{[A(k_p)]^2}{(\hbar \omega_{\rm M}+\varepsilon_h) 
{(\sqrt{\hbar \omega_{\rm M}+\varepsilon_h}-\sqrt{\varepsilon_h})}^4 
{(\sqrt{\varepsilon_p}-\sqrt{\hbar \omega_{\rm M}+\varepsilon_h})}^2}  ,
\end{align*}
with $\varepsilon_h = \varepsilon_p-2 \hbar \omega_{\rm M}$.

Taking the limit of $L \rightarrow \infty$, we finally obtain
\begin{equation}
\label{Gamma_ion}
\Gamma_{\rm ion}(a) \simeq \frac{3 \pi}{80} \frac{\varepsilon_{\rm F}}{k_{\rm F}a} q(\xi, \zeta)  ,
\end{equation}
where $\xi = \hbar \omega_{\rm M}/\varepsilon_{\rm F}$ and $\zeta = W/\varepsilon_{\rm F}$. The 
function $q$ of the two variables $\xi$ and $\zeta$ is defined in Appendix \ref{sec_gh}, Eq.~\eqref{q}.

The size scaling of $\Gamma_{\rm ion}$ is mainly given by a $1/a$-dependence of the prefactor
(Fig.~\ref{fig_gamma_ion}), despite the fact that the work function W appearing in the parameter
$\zeta$ is size dependent and scales (for a neutral cluster) as \cite{seidl}
$W = W_{\infty}+ 3e^2/8a$ where $W_{\infty}$ is the work function of the bulk material.

Using the work function $W = 4.65$ eV and the experimental value of $\hbar \omega_{\rm M} = 2.75$ eV 
for the charged Na$_{93}^+$ clusters of Ref.~\onlinecite{schlipper}, Eq.~\eqref{Gamma_ion} yields 
$\Gamma_{\rm ion} \simeq 0.1$ eV, which corresponds to an ionization lifetime of the second plasmon 
of $6.6$ fs. This value is of the same order of magnitude as the experimentally reported lifetime of 
$10$ fs. It is also in rough agreement with the estimation yielded by the numerical calculations of
Ref.~\onlinecite{bertsch:dp} based on a separable residual interaction ($10$ to $20$ fs). Therefore, 
despite the approximations we have been forced to make in our analytical calculations, we believe that we 
kept the essential ingredients of this complicated problem. The lifetimes obtained by the different 
procedures consistently establish the second plasmon as a well-defined resonance in metallic clusters. 
While the numerical calculations of Ref.~\onlinecite{bertsch:dp} have been performed for just one size, 
our results exhibit a clear size dependence that can be tested in future experiments.

%=============================================================================
%=============================================================================
%=============================================================================
% CONCLUSION
\section{Conclusion}
\label{sec_ccl}
In this work we have analyzed the lifetime of collective excitations in metallic clusters. Different decay
mechanisms have been studied within a semiclassical approach for the mean-field self-consistent 
potential describing the electrons in a jellium background. We have considered Landau damping, which is
the dominant relaxation mechanism for nanoparticles with radius $a$ in the range 0.5--2.5 nm. We
found that the linewidth of the single surface plasmon exhibits a $1/a$ dependence, superimposed 
to an oscillating behavior arising from electron-hole density-density correlations. These results are in
good agreement with numerical time-dependent local density approximation calculations,  and consistent with
experiments on free alkaline nanoparticles. 

To describe noble metal clusters, we have taken into account the screening effect of the d electrons and the
modifications induced by the dielectric properties of an eventual matrix. We have demonstrated that such 
an inhomogeneous dielectric environment of the nanoparticles strongly affects the steepness of the 
self-consistent potential, which in turn has a crucial influence on the plasmon linewidth. We could then 
solve the discrepancy presented in Ref.~\onlinecite{molina} between the well-known Kawabata and Kubo 
formula on one  side, against experiments and numerical calculations on the other side. The
size-dependent oscillations of the linewidth also depend on the dielectric constants through the slope of 
the self-consistent potential. The access to individual nano-objects, recently developped by different
experimental techniques, provides a promising way of testing our theoretical results concerning the 
size-dependent linewidth oscillations. 

The physical relevance of the second plasmon has been analyzed in terms of different decay channels:
Landau damping and particle ionization. We have shown that both processes are relevant, but they do
not preclude the existence of the resonance. The comparison of our semiclassical calculation with
the existing numerical and ex\-pe\-ri\-men\-tal results is reasonably good, despite the various approximations of our model. 

Our theoretical results concerning the different decay mechanisms of the collective excitations of 
metallic clusters should be important for the analysis of the electron dynamics following short and strong 
laser excitations.

%=============================================================================
%=============================================================================
% ACKNOWLEDGMENTS
\begin{acknowledgments}
We are grateful to G.-L.~Ingold for his careful reading of the manuscript and for illuminating suggestions.
We thank G.~F.~Bertsch, J.-Y.~Bigot, and P.-A. Hervieux for useful discussions. We acknowledge the Centre de Coop\'eration
Universitaire Franco-Bavarois/Bayerisch-Franz\"osisches
Hochschulzentrum (CCUFB/BFHZ), the French-German PAI PROCOPE, and the EU (RTN program) 
for financial support. 
\end{acknowledgments}

%=============================================================================
%=============================================================================
%=============================================================================
%=============================================================================
%=============================================================================
%=============================================================================
%=============================================================================
%=============================================================================
%=============================================================================
% APPENDIX
\appendix

%=============================================================================
%=============================================================================
%=============================================================================
%=============================================================================
% TRANSITION POTENTIAL
\section{Transition potential}
\label{mismatch}
In this Appendix we present the derivation of the transition potential induced by the plasmon field and
ge\-ne\-ra\-li\-ze the derivation of Ref.~\onlinecite{bertsch} considering the Coulomb interaction in the 
case of a dielectric mismatch between the electrons and the surrounding matrix in which the nanoparticles 
are embedded. Assuming that at e\-qui\-li\-bri\-um the electron density is uniform within a sphere of 
radius $a$, $n({\bf r}) = n \Theta(a-r)$ ($\Theta$ being the Heaviside distribution), a rigid displacement 
with a magnitude $Z$ along the ${\bf e}_z$ direction changes the density at $\bf r$ from $n({\bf r})$ to
\begin{equation*}
n({\bf r}-{\bf u}) = n({\bf r}) + \delta n({\bf r})  .
\end{equation*}
To first order in the field ${\bf u} = Z {\bf e}_z$, we can write
\begin{equation*}
\delta n({\bf r}) = -{\bf u}\cdot \frac{\partial}{\partial {\bf r}} n({\bf r})
= Z n \cos{\theta} \delta(r-a)  .
\end{equation*}
We have neglected the oscillations of the density in the inner part of the particle due to 
shell effects, and also the extension of the electronic density outside of the particle (spillout effect). 
\cite{deheer} Noting $V_{\rm C}({\bf r},{\bf r'})$ the Coulomb electron-electron interaction, the change in 
the self-consistent potential due to the rigid shift (transition potential) is
\begin{equation}
\label{transition_potential}
\delta V({\bf r}) = \int {\rm d}^3{\bf r'} \delta n({\bf r'}) V_{\rm C}({\bf r},{\bf r'})  .
\end{equation}
Using the multipolar decomposition of the Coulomb interaction, one obtains \cite{bertsch}
\begin{equation}
\label{deltaV}
\delta V({\bf r}) = Z \frac{4\pi n e^2}{3} d({\bf r})  ,
\end{equation}
with
\begin{equation}
\label{dipole_field}
d({\bf r}) = 	
\begin{cases}
	\displaystyle z   , &r \leqslant a  , \\
	\displaystyle  \frac{za^3}{r^3}  , & r > a  .
\end{cases}
\end{equation}
We notice that a displacement of the electron system leads to a dipolar field inside the nanoparticle, and 
that its magnitude decays as $1/r^2$ outside the particle.

If we now consider the case of a noble metal nanoparticle (where the d electrons are taken into account
with the help of a dielectric constant $\epsilon_{\rm d}$) embedded in a matrix (of dielectric constant 
 $\epsilon_{\rm m}$), the Coulomb interaction between electrons is given by \cite{serra}
\begin{widetext}
\begin{equation*}
V_{\rm C}({\bf r},{\bf r'}) = 4 \pi e^2 
\begin{cases}
	\displaystyle  
  	\frac{1}{\epsilon_{\rm d}} \sum_{lm} \frac{1}{2l+1} \left[ \frac{r_<^l}{r_>^{l+1}}
 	+ \frac{r^{l} r'^{l}}{a^{2l+1}} \frac{(l+1)(\epsilon_{\rm d}-\epsilon_{\rm m})}
 	{\epsilon_{\rm d} l+\epsilon_{\rm m}(l+1)} \right]
 	{Y_l^m}^\ast (\Omega) Y_l^m(\Omega')  ,
  	& r, \ r' \leqslant a , \\
	\displaystyle 
  	 \sum_{lm} \frac{r_<^l}{r_>^{l+1}} \frac{{Y_l^m}^\ast (\Omega) Y_l^m(\Omega')}
 	{\epsilon_{\rm d} l+\epsilon_{\rm m}(l+1)}  ,
  	& r_<\leqslant a, \ r_>>a ,  \\
	\displaystyle 
	\frac{1}{\epsilon_{\rm m}} \sum_{lm} \frac{1}{2l+1} \left[ \frac{r_<^l}{r_>^{l+1}} 
 	+ \frac{a^{2l+1}}{r^{l+1} r'^{l+1}} \frac{l(\epsilon_{\rm m}-\epsilon_{\rm d})}
 	{\epsilon_{\rm d} l+\epsilon_{\rm m}(l+1)} \right]
 	{Y_l^m}^\ast (\Omega) Y_l^m(\Omega')  ,
  	& r, \ r' > a  ,
\end{cases}
\end{equation*}
\end{widetext}
where $r_< = \min{(r, r')}$, $r_> = \max{(r,r')}$, and $Y_l^m$ are the spherical harmonics. Inserting
this expression into Eq.~\eqref{transition_potential}, we obtain the result of Eq.~\eqref{deltaV}  
with the additional multiplying factor $3/(\epsilon_{\rm d}+2 \epsilon_{\rm m})$. 

In both cases (with and without a dielectric mismatch), the expression \eqref{deltaV} can be written 
as $\delta V({\bf r}) = Z m_{\rm e} \omega_{\rm M}^2 d({\bf r})$. The only effect of the dielectric constants 
on the transition potential as compared to the free case is through the red-shift of the Mie frequency.

%=============================================================================
%=============================================================================
%=============================================================================
%=============================================================================
% SEMICLASSICS WITH RADIAL SYMMETRY
\section{Semiclassics with radial symmetry}
\label{sec:semiclassics}
Semiclassical expansions constitute a very useful tool in mesoscopic physics since they allow for an 
intuitive description of relatively complex systems. The spectral properties of metallic clusters 
\cite{brack} or the conductance fluctuations in the electronic transport through quantum dots 
\cite{jalabert_review} can be readily understood when the quantum observables are expressed in 
terms of an appropriate ensemble of classical trajectories.

In problems with radial symmetry, like the one we treat in this work, it is tempting to take advantage of the 
se\-pa\-ra\-bi\-li\-ty into radial and angular coordinates in order to reduce the dimensionality of the
trajectories con\-tri\-bu\-ting to the semiclassical expansions. However there are technical difficulties introduced by the 
singularity at the origin of the centrifugal potential, and this is probably the reason why the radial
symmetry is often not fully exploited in semiclassical expansions. On the other hand, the well-known 
Langer modification \cite{langer} is a prescription to avoid the above-mentioned difficulties and provides 
a route to the semiclassical quantization of spherically symmetric systems (which has been recently 
extended to higher orders \cite{grabert}).

In this Appendix we start from the Langer modification in order to obtain the partial (or
angular momentum dependent) density of states (DOS) $\varrho_l(\varepsilon)$ that we need in our 
evaluation of plasmon lifetimes. As a check of consistency, we verify in a few simple examples that when 
$\varrho_l(\varepsilon)$ is summed (in a semiclassical way) over $l$ and $m$, we recover the well-known 
Berry-Tabor formula for the total DOS. \cite{berry,balian}

%=============================================================================
%=============================================================================
%=============================================================================
% LANGER MODIFICATION AND PARTIAL DOS
\subsection{Langer modification and partial density of states}
\label{app_langer}
For a central potential $V(r)$, the Schr\"odinger equation is separable into angular and radial parts.
The wave function can be written as $\psi_{klm}({\bf r}) = [u_{kl}(r)/r] Y_l^m(\Omega)$, where $u_{kl}$ 
verifies
\begin{equation}
\label{1DSch}
\left[ -\frac{\hbar^2}{2m_{\rm e}} \frac{{\rm d}^2}{{\rm d}r^2} + \frac{\hbar^2 l(l+1)}{2m_{\rm e} r^2} +V(r) \right] u_{kl}(r)
= \varepsilon_{kl} u_{kl}(r)  ,
\end{equation}
with the condition $u_{kl}(0) = 0$. It is important to notice that the variable $r$ is limited to positive values 
and that the centrifugal potential possesses a singularity at $r=0$. This significant difference between 
Eq.~\eqref{1DSch} and a standard one-dimensional Schr\"odinger equation prevents from a naive
application of the Wentzel-Kramers-Brillouin (WKB) approximation to treat this radial pro\-blem. The 
change of variables $x = \ln{r}$ and $\chi_{kl}(x) = \exp{(x/2)} u_{kl}(r)$ results in a standard  Schr\"odinger 
equation. Using the WKB approximation for $\chi_{kl}$ amounts to change the centrifugal potential in 
Eq.~\eqref{1DSch} according to the Langer modification \cite{langer,mount}
\begin{equation*}
l(l+1) \Longrightarrow \left( l+ \frac 12 \right)^2  .
\end{equation*}
The resulting WKB quantization provides the exact spectrum for the hydrogen atom, as well as for
the three-dimensional isotropic harmonic oscillator.

The same kind of considerations in two-dimensional systems with a circular symmetry leads to the 
following substitution in the centrifugal potential: \cite{brack,mount}
\begin{equation}
\label{LM2D}
\left( m-\frac 14 \right)^2 \Longrightarrow m^2  ,
\end{equation}
which yields an exact WKB spectrum for the cases of the isotropic harmonic oscillator as well as for the 
hydrogen atom in two dimensions.

The semiclassical approximation provides a method to calculate the leading $\hbar$ contributions to the 
DOS in the limit of large quantum numbers, and decomposes the DOS into a smooth and an oscillating
part. The smooth term is simply the Weyl contribution \cite{brack} and the  oscillating term is given, in the 
case where the periodic orbits (POs) are not degenerated in action, by the Gutzwiller trace formula
\cite{gutzwiller} as a sum over the primitive periodic orbits (PPOs).

In the case of multidimensional integrable systems, the POs belonging to a torus of the phase space are
degenerate, and the oscillating part of the DOS is given by the Berry-Tabor formula as a sum over 
rational tori. \cite{berry} In one-dimensional problems, or in the radial coordinate of a spherically
symmetric case, the trajectories are not degenerate, and therefore the semiclassical approximation to the
DOS at fixed angular momentum $l$ is given by
\begin{equation}
\label{trace}
\varrho_l(\varepsilon) = \varrho_l^0(\varepsilon) + \varrho_l^{\rm osc}(\varepsilon)   ,
\end{equation}
with 
\begin{align*}
\varrho_l^0(\varepsilon) &= \frac{\tau_l(\varepsilon)}{2 \pi \hbar}  ,\\
\varrho_l^{\rm osc}(\varepsilon) & =  \frac{\tau_l(\varepsilon)}{ \pi \hbar}
 \sum_{\tilde r=1}^{\infty} 
 \cos{\left[ \tilde r \left( \frac{S_l(\varepsilon)}{\hbar}
 -\nu_{\rm c} \frac{\pi}{2}-\nu_{\rm r} \pi \right) \right]}  ,
 \end{align*}
where $S_l$ and $\tau_l = \partial S_l/\partial \varepsilon$ are the action and period refering to the 
motion in the effective ($l$-dependent) radial potential; $\nu_{\rm c}$ ($\nu_{\rm r}$) is the number of
classical turning points of the PPOs against the smooth (hard) walls.

%=============================================================================
%=============================================================================
%=============================================================================
% TOTAL DOS: BERRY-TABOR DOS WITH RADIAL SYMMETRY
\subsection{Total density of states and Berry-Tabor formula for systems with radial symmetry}
\label{sec:berry_tabor}
Using the selection rules for the plasmon decay, its lifetime can be expressed in terms of the partial DOS  
$\varrho_l(\varepsilon)$ whose semiclassical expression is given by Eq.~\eqref{trace}. It is then important to
 verify that the semiclassical sum over angular momenta (that we use throughout our calculations),
when applied to $\varrho_l(\varepsilon)$, is able to reproduce the total DOS. Rather than working the most  
general case, we perform our test for three particular examples: the disk billiard (where the calculations are 
particularly simple), the three dimensional billiard (like the one we treat in the text), and the isotropic 
spherical harmonic oscillator (where the semiclassical spectrum coincides with the exact one).

%=============================================================================
%=============================================================================
% DISK BILLIARD 
\subsubsection{Disk billiard}
\label{sec:disk}
A disk billiard is defined by its radial potential
\begin{equation}
\label{radialpot}
V(r) = 
\begin{cases}
	0  ,& r < a , \\
	\infty  , & r \geqslant a  ,
\end{cases}
\end{equation} 
where $a$ is the radius of the disk. The effective radial motion is governed by the potential
$V_m^{\rm eff}(r) = \hbar^2 m^2/2 m_{\rm e} r^2+V(r)$, with $m$ the $z$ component of the angular
momentum included according to Eq.~(\ref{LM2D}). The classical PPOs have 
$\nu_{\rm c} = \nu_{\rm r} = 1$
since there is one turning point at the (smooth) kinetic barrier and another at the (hard) wall for $r=a$.
For a given energy $\varepsilon$ we have 
$m_{\rm max} = (2m_{\rm e} \varepsilon)^{1/2}a/\hbar = (\varepsilon/\varepsilon_0)^{1/2} = ka$, with 
$\varepsilon_0 = \hbar^2/2m_{\rm e} a^2$. 

The action and period of the PO with energy $\varepsilon$ and angular momentum $m$ are given by
\begin{subequations}
\label{m}
\begin{align}
S_m(\varepsilon) &= 2 \hbar \left[ \sqrt{(ka)^2-m^2}-m \arccos{\left(  \frac{m}{ka} \right)} \right]  , \\
\tau_m(\varepsilon) &= \frac{\hbar \sqrt{(ka)^2-m^2}}{\varepsilon_0 (ka)^2}  ,
\end{align}
\end{subequations}
respectively. The smooth part of the DOS is
\begin{equation*}
\varrho^0(\varepsilon) = \sum_{m = - m_{\rm max}}^{+m_{\rm max}} \varrho_{m}^0(\varepsilon) 
= \frac{1}{4 \pi} \left( \frac{2 m_{\rm e}}{\hbar^2} \right) {\cal A}   ,
\end{equation*}
with ${\cal A} = \pi a^2$ being the disk area. We have replaced the sum by an integral and obtained the 
Weyl part of the DOS. For the oscillating part we make use of the Poisson summation rule and write
\begin{equation*}
\varrho^{\rm osc}(\varepsilon) = \sum_{\tilde m=-\infty}^{+\infty} 
\sum_{\substack{\tilde r\geqslant1\\ \sigma = \pm}}
\int_0^{m_{\rm max}} {\rm d}m \frac{\tau_m(\varepsilon)}{2 \pi \hbar} \ {\rm e}^{\sigma {\rm i} 
\phi_m^{\tilde m \tilde r}(\varepsilon)} 
\end{equation*}
with the phase
\begin{equation*}
\phi_m^{\tilde m \tilde r}(\varepsilon) = \tilde r \left[ \frac{S_m(\varepsilon)}{\hbar}-\frac{3 \pi}{2} \right] + 2\pi \tilde m m  .
\end{equation*}
Consistently with the semiclassical expansions, we perform a stationary phase approximation. The 
stationary points are given by 
$\bar{m} = ka \cos{\varphi_{\tilde r \tilde m}}$, with $\varphi_{\tilde r \tilde m} = \pi \tilde m/\tilde r$ and 
the condition $\tilde r \geqslant 2 \tilde m > 0$, which yield just the classical angular momenta of the POs 
labeled by the topological indices $(\tilde r,\tilde m)$. We then recover for the oscillating DOS the 
well-known result \cite{berry,balian}
\begin{equation*}
\varrho^{\rm osc}(\varepsilon) = \frac{1}{\varepsilon_0}\frac{1}{\sqrt{\pi ka}} \sum_{\tilde m=1}^{\infty}
\sum_{\tilde r\geq 2\tilde m} f_{\tilde r \tilde m} \frac{\sin{^{3/2}\varphi_{\tilde r \tilde m} }}
{\sqrt{\tilde r}}  \cos{\Phi_{\tilde r \tilde m}}  ,
\end{equation*}
where $f_{\tilde r \tilde m}=1$ if $\tilde r=2\tilde m$ and $f_{\tilde r \tilde m}=2$ if $\tilde r>2\tilde m$, 
$\Phi_{\tilde r \tilde m} = kL_{\tilde r \tilde m} -3r\pi/2+\pi/4$ and 
$L_{\tilde r \tilde m} = 2 \tilde ra \sin{\varphi_{\tilde r \tilde m}}$ is the length of the orbit 
$(\tilde r, \tilde m)$.

We also notice that the quantization of the radial problem leads to the well-known Keller and Rubinow
condition \cite{keller}
\begin{equation*}
\sqrt{(ka)^2-m^2}-m \arccos{\left( \frac{m}{ka} \right)} = \pi \left(n+\frac 34\right)  ,
\end{equation*}
from which the Berry-Tabor formula can be readily obtained.

%=============================================================================
%=============================================================================
% SPHERICAL BILLIARD 
\subsubsection{Spherical billiard}
\label{sec:berry_tabor_sphere} 
A spherical billiard is also defined by Eq.~(\ref{radialpot}), but in the three-dimensional case
$V_l^{\rm eff}(r) = \hbar^2(l+1/2)^2/2m_{\rm e} r^2 + V(r)$. For an energy $\varepsilon$, the maximum value of 
the angular momentum is given by $l_{\rm max} = ka-1/2 $. The action 
and the period of a trajectory with energy $\varepsilon$ and angular momentum $l$ are the same as in the 
two-dimensional case up to a change of $m$ by $l+1/2$ [Eqs.~\eqref{m}].
The total DOS is given by
\begin{equation*}
\varrho(\varepsilon) = \sum_{l=0}^{l_{\rm max}} \sum_{m = -l}^{+l} \varrho_l(\varepsilon) 
= \sum_{l=0}^{l_{\rm max}} (2l+1) \varrho_l(\varepsilon)  .
\end{equation*}
For the smooth DOS, we find the first term of the Weyl expansion: \cite{brack}
\begin{equation*}
\varrho^0(\varepsilon) = \frac{1}{4 \pi^2} \left( \frac{2m_{\rm e}}{\hbar^2} \right)^{3/2} 
\sqrt{\varepsilon} \ {\cal V}  ,
\end{equation*}
where ${\cal V} = 4 \pi a^3/3$ is the volume of the sphere, and for the oscillating part
\begin{align*}
&\varrho^{\rm osc}(\varepsilon) = \frac{1}{\varepsilon_0} \sqrt{\frac{ka}{\pi}} 
 \\
&\times
\sum_{\substack{ \tilde m \geqslant 1 \\ \tilde r>2\tilde m }} 
(-1)^{\tilde m} \sin{(2 \varphi_{\tilde r \tilde m})}
\sqrt{\frac{\sin{\varphi_{\tilde r \tilde m}}}{\tilde r}} \cos{\Phi_{\tilde r \tilde m}}  ,
\nonumber
\end{align*}
with the same notations as in Appendix \ref{sec:disk}. We again recover the Berry and Tabor
semiclassical DOS to leading order in $\hbar$, \cite{berry,balian} as well as the quantization condition of
Keller and Rubinow. \cite{keller}

%=============================================================================
%=============================================================================
% 3D HARMONIC OSCILLATOR 
\subsubsection{Isotropic spherical harmonic oscillator}
The isotropic harmonic oscillator in three dimensions is a nonbilliard integrable system and therefore
the Berry-Tabor quantization is very difficult to implement. The radial approach that we develop
clearly overcomes this difficulty. The effective potential is 
$V_l^{\rm eff}(r) = \hbar^2(l+1/2)^2/2m_{\rm e} r^2+ (1/2) m_{\rm e} \omega^2r^2$,
where $\omega$ is the pulsation of the harmonic confinement. At a given $\varepsilon$, we have
$l_{\rm max} = -1/2+\varepsilon/\hbar \omega$. The classical action is given by
$S_l(\varepsilon) = \varepsilon \pi/\omega - \pi \hbar (l+1/2)$ and the period is $\tau = \pi/\omega$. 
Using Eq.~(\ref{trace}) with $\nu_{\rm c} = 2r$ and $\nu_{\rm r}=0$ (no hard wall) gives the DOS at
fixed orbital momentum.

For the smooth part of the DOS, the sum over $l$ can be performed exactly, but to be consistent with the 
semiclassical approximation we have to take the limit $\varepsilon/\hbar \omega \gg 1$: 
$\varrho^0(\varepsilon) \simeq \varepsilon^2/2(\hbar \omega)^3$.
Writing the Poisson summation rule for the oscillating part and performing a stationary phase 
approximation, we have the condition on topological indices $\tilde r = 2\tilde m$ and 
$\tilde m \geqslant 1$. Finally we obtain for the total DOS the trace formula
\begin{equation*}
\varrho(\varepsilon) = \frac{\varepsilon^2}{2(\hbar \omega)^3} \left[ 1+2 \sum_{\tilde m=1}^\infty 
(-1)^{\tilde m} \cos{\left( 2\pi \tilde m \frac{\varepsilon}{\hbar \omega} \right)} \right]  ,
\end{equation*}
which has to be compared with the exact trace formula given in Ref.~\onlinecite{brack}, where the 
prefactor is shifted by the quantity $-1/8 \hbar \omega$, negligible at the (high energy) semiclassical 
limit. One also notices that the WKB quantization rule yields the exact quantum spectrum of the harmonic 
oscillator: $\varepsilon_{nl} = \hbar \omega (2n+l+3/2)$. \\

We have demonstrated the usefulness of the radial decomposition for the semiclassical expansion of the 
DOS. Even in the case of degenerated classical periodic trajectories, one is able to find the semiclassical
DOS by using the appropriate symmetry of the system, without requiring the action-angle quantization of 
Berry and Tabor.

%=============================================================================
%=============================================================================
%=============================================================================
% DIPOL MATRIX ELEMENT WITH RADIAL SYMMETRY
\subsection{Semiclassical dipole matrix element with spherical symmetry}
\label{sec:dipol_ME}
In this Appendix we focus on the semiclassical evaluation of the dipole matrix element for the case of
a sphe\-ri\-cal\-ly symmetric system, and extend the well-known result which relates in the 
one-dimensional case the dipole matrix element to the Fourier components of the classical motion of the 
particle.\cite{landau}

The spherical symmetry permits us to separate the dipole matrix element $\langle nlm | z | n'l'm' \rangle$ 
into two parts: an angular part given by Eq.~\eqref{sp_angular} and a radial part 
\begin{equation*}
{\cal R}_{nn'}^{ll'} = \frac{\hbar^2}{m_{\rm e} (\varepsilon_{n'l'}-\varepsilon_{nl})} 
\int_0^\infty {\rm d}r \ u_{nl}(r)  \frac{{\rm d}}{{\rm d}r} u_{n'l'}(r)  ,
\end{equation*}
where the radial wave functions $u_{nl}$ satisfy Eq.~\eqref{1DSch} and we have used the commutation 
relation between the radial momentum and the Hamiltonian. Next we restrict ourselves to the classical 
region in the effective potential $V_l^{\rm eff}(r)$ between the two turning points $(r_-,r_+)$ and use the 
WKB approximation to express the radial wave functions as
\begin{equation*}
u_{nl}(r) = 
\frac{2 \cos{\displaystyle \left\{ 1 /\hbar \int_{r_-}^r {\rm d}r' \sqrt{2m_{\rm e} \left[\varepsilon_{nl}-V^{\rm eff}_l(r')\right]}-\pi/ 4
\right\}}}{\sqrt{\tau_l} \left\{2m_{\rm e} \left[\varepsilon_{nl}-V^{\rm eff}_l(r)\right]/m_{\rm e}^2 \right\}^{1/4}}
 .
\end{equation*}
We also assume that the radial potential is a smoothly varying function of the radial coordinate, that 
$l \simeq l'$ (this is justified because the selection rules dictate that $l' = l \pm 1$ and we are in the high
energy limit) and that the energies involved in the dipole matrix element are sufficiently close to each 
other to satisfy
\begin{equation}
\label{energy_diff}
\varepsilon_{n'l'}-\varepsilon_{nl} \approx \frac{2 \pi \hbar \Delta n}{\tau_l}  ,
\end{equation}
with $\Delta n = n'-n$.

With these approximations, changing the spatial coordinate $r$ to the time $t$, we obtain, to leading 
order in $\hbar$ 
\begin{equation}
\label{dipol_ME_sc}
{\cal R}_{nn'}^{ll'} = \frac{2}{\tau_l} 
\int_0^{\tau_l/2} {\rm d}t \ r(t) \cos{\left( 2 \pi \Delta n \frac{t}{\tau_l} \right)}  ,
\end{equation}
where $r(t)$ represents the classical trajectory in the effective potential. Thus we see that, as in the 
one-dimensional case, the dipole matrix element is related to the Fourier transform of the trajectory of the 
classical motion. 

As a check of consistency, we apply this semiclassical analysis to the hard-wall potential involved in 
our eva\-lu\-ation of the surface-plasmon lifetime. This analysis is only possible in the limit
$\varepsilon_{\rm F} \gg \hbar \omega_{\rm M}$: The approximation of
Eq.~\eqref{dipol_ME_sc} is valid if we assume that the energy of the particle is close to the 
one of the hole. This energy dif\-fe\-rence is, because of the conservation of energy appearing in the 
Fermi Golden Rule \eqref{Gamma_sp}, simply $\hbar \omega_{\rm M}$. 

At a given energy $\varepsilon$, the periodic trajectory in the effective potential is 
\begin{equation*}
r(t) = \sqrt{\frac{2\varepsilon}{m_{\rm e}}t^2+\frac{\hbar^2(l+1/2)^2}{2 m_{\rm e} \varepsilon}}
\ , \ 0 \leqslant t \leqslant \frac{\tau_l}{2}  .
\end{equation*}
Substituting this expression in Eq.~\eqref{dipol_ME_sc} and making the expansion in $1/\Delta n$
[proportional to $1/\hbar \omega_{\rm M}$, see Eq.~\eqref{energy_diff}], we obtain the leading order term
\begin{equation}
\label{dipol_ME_sc_infiniteslope}
{\cal R}_{k_p k_h}^{l_p l_h} = \frac{2 \hbar^2}{m_{\rm e} a} \frac{\varepsilon_p}
{{(\varepsilon_p-\varepsilon_h)}^2}  ,
\end{equation}
which agrees with Eq.~\eqref{sp_radial} in the limit $\varepsilon_p \approx \varepsilon_h$. We notice 
that this semiclassical dipole matrix element leads to the correct result for the smooth part $\Gamma^0$ 
of the single-plasmon linewidth in the limit 
$\xi = \hbar \omega_{\rm M}/\varepsilon_{\rm F} \rightarrow 0$ of Eq.~\eqref{sp_smooth}.

%=============================================================================
%=============================================================================
%=============================================================================
%=============================================================================
% W & H
%\begin{widetext}
\section{Frequency dependence of the plasmon linewidths}
\label{sec_gh}
\begin{figure}[h]
\begin{center}
\includegraphics[width=8truecm]{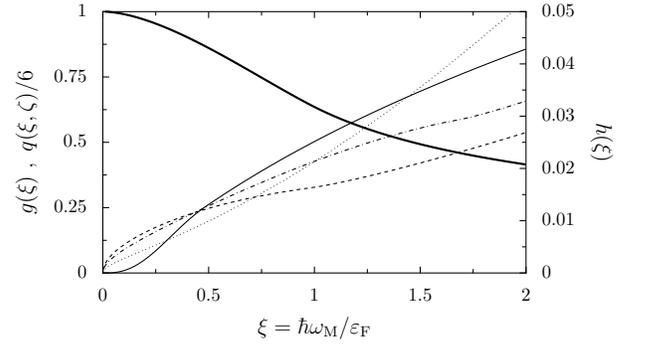}
\caption{\label{functions} 
Functions $g(\xi)$ (thick line), $h(\xi)$ (full line), and $q(\xi, \zeta)$.
The function $q(\xi, \zeta)$ is represented as a function of $\xi$ for $\zeta/\xi = 1$ (dashed line),
$1.4$ (dashed-dotted line), and $1.8$ (dotted line). }
\end{center}
\end{figure}

In this Appendix, we present the frequency dependence of the single- and double-plasmon linewidths.
In Fig.~\ref{functions}, we represent the function $g$ (thick line) of
$\xi = \hbar \omega_{\rm M}/\varepsilon_{\rm F}$ involved in the expression of the single-plasmon 
linewidth [see Eq.~\eqref{sp_smooth}], as well as in the first-order decay rate of the double plasmon
($\Gamma_{2 \rightarrow 1}$). The function $g$ is plotted after its analytical expression [Eqs.~(62) and
(63) in Ref.~\onlinecite{yannouleas}] and is a smoothly decreasing function with $\lim_{\xi \rightarrow \infty} g(\xi)= 0$.

The function $h$ involved in the expression of the second-order double-plasmon linewidth 
\eqref{2P_academic} is defined by
\begin{align*}
h(\xi) =  \int_{\max{(1, 2 \xi)}}^{1+2\xi} 
&{\rm d}z \int_0^{z-2\xi} {\rm d}y \sqrt{z-y} \sqrt{z-y-2\xi} \nonumber \\
&\times
\left( \sqrt{\frac{z-y}{z}}-\sqrt{\frac{z-y-2 \xi}{z-2\xi}} \right)^2 
\end{align*}
and has been approximately determined by integrating out the intermediate states $i$ in the limit 
$k_{\rm F}a \gg 1$. The integral over the intermediate state energy has been performed by introducing 
cutoffs in order to avoid unphysical divergencies due to the fact that discrete single-particle levels have
been replaced in our model by a continuum of states. When the remaining two-dimensional integral is evaluated
numerically we obtain a smoothly increasing function of the parameter $\xi$, with $h(0) = 0$ as
shown in Fig.~\ref{functions}. This function has the asymptotic limit
$\lim_{\xi \rightarrow \infty} h(\xi) = \infty$. We see that when the double-plasmon state is too high in 
energy, the linewidth diverges to infinity and this resonance is no longer well-defined (the 
double-plasmon state has a lifetime equal to zero in this condition). 

The function $q$ of the two variables $\xi$ and $\zeta = W/\varepsilon_{\rm F}$ involved in our 
evaluation of the ionization rate via the double-plasmon state, Eq.~\eqref{Gamma_ion}, is defined as 
\begin{align}
\label{q}
q(\xi, \zeta) &= \left( \frac{\xi}{2} \right)^6 
\int_{\max{(2 \xi, 1+\zeta)}}^{1+2 \xi} {\rm d}z
\frac{(2z-1-\zeta) \sqrt{z-2\xi}}{z \sqrt{(z-\xi)(z-1-\zeta)}} \nonumber \\
&\times
\frac{1}{{(\sqrt{z-\xi}-\sqrt{z-2\xi})}^4{(\sqrt{z}-\sqrt{z-\xi})}^2}  .
\end{align}
Since our approach is valid when $\hbar \omega_{\rm M} \leqslant W \leqslant 2 \hbar \omega_{\rm M}$, the function
$q$ is defined for $\xi \leqslant \zeta \leqslant 2 \xi$ and can be integrated numerically.
The result is shown in Fig.~\ref{functions}. The function $q$ is not very sensitive to the value of
$\zeta/\xi = W/\hbar \omega_{\rm M}$ in the presented interval. However, it vanishes at the upper limit 
($W = 2 \hbar \omega_{\rm M}$), since in this case particle states cannot be in the continuum and 
$\Gamma_{\rm ion} = 0$.

%=============================================================================
%=============================================================================
%=============================================================================
%=============================================================================
%=============================================================================
%=============================================================================
% BIBLIOGRAPHY


\begin{thebibliography}{}
\bibitem{wavelength}
This is fulfilled for the nanoparticles we consider. The plasmon frequency is in the optical range and the 
corresponding wavelength is much larger than $a$.
\bibitem{footnote_Na}
The discrepancy in Fig.~\ref{fig_phase_diag_Ag}(b) for $\epsilon_{\rm d} = \epsilon_{\rm m} = 1$ between
Eq.~\eqref{sp_smooth} and the TDLDA is due to the uncertainty of the fitting procedure. For 
$r_{\rm s} = 3.93$ $a_0$ (not shown), both points are in good agreement.
\bibitem{footnote}
As a test of consistency, we checked that this approximation leads in the case of the single plasmon to a 
linewidth given by Eq.~\eqref{sp_smooth} for $\xi = 0$. For $\xi \neq 0$, we obtain a 
slightly different frequency-dependence, but with the same decreasing feature.
\end{thebibliography}
\end{document}